\documentclass{aa}
\usepackage{longtable}
\usepackage{lscape}
\usepackage{latexsym}
\usepackage{amssymb}

\usepackage[dvips]{graphicx}
\input{psfig.txt}

\begin{document}

\title{Spectrophotometry of galaxies in the Virgo cluster.II: The data 
\thanks{Based 
on observations collected at the  Observatoire de Haute Provence (OHP) (France), 
operated by the CNRS, at the
European Southern Observatory (proposals 66.B-0026; 68.B-0505), at the Loiano telescope belonging 
to the University of Bologna (Italy) and at the Observatorio Astronomico National de San Pedro Martir (Mexico).}
}

\subtitle{}

\author{Giuseppe Gavazzi\inst{1}
\and Antonio Zaccardo\inst{1}
\and Gerry Sanvito\inst{1}
\and Alessandro Boselli\inst{2}
\and Christian Bonfanti\inst{1}
}

\offprints{G. Gavazzi}

\institute{
Universit\`a degli Studi di Milano - Bicocca, P.zza delle scienze 3,
20126 Milano, Italy.
\and
Laboratoire d'Astrophysique de Marseille, Traverse du Siphon, F-13276 Marseille
Cedex 12, France.
}

\date{Received July 20, 2003; accepted..........}

\abstract{
Drift-scan mode (3600-6800 \AA) spectra with $500<R<1000$ resolution  are presented 
for 333 galaxies members to nearby clusters, covering the whole Hubble sequence. The majority (225) were obtained 
for galaxies in the Virgo cluster where a completeness of 36 \%, if all Hubble types are considered, and of 51 \%, 
restricting to late-types, was reached at $m_p\leq$16.  Our data
can be therefore considered representative of the integrated spectral 
properties of giant and dwarf galaxies in this cluster. Intensities and 
equivalent widths (EWs) are derived for the principal lines, both in emission and 
in absorption. Deblending of the underlying absorption from emission was achieved in most cases.
}
\titlerunning{Spectro-photometry of galaxies in the Virgo cluster}
\authorrunning{G. Gavazzi, et al.}
\maketitle
\section {Introduction}

Local galaxies are the relics of evolutionary processes that took place in the universe since the early 
collapse of primordial matter fluctuations up to the present cosmological epoch. Such processes have not 
yet been convincingly unveiled, in spite of an increasing observational effort involving 
today's major observational facilities.
A satisfactory characterization of the properties of local galaxies is building up only recently, 
as data obtained through a variety of observational windows of the electromagnetic spectrum are being gathered
(see Kennicutt, 1998; Roberts \& Haynes, 1994; Gavazzi, Pierini \& Boselli, 1996). 
Complete imaging data sets taken in a broad frequency range are becoming available owing to  
extensive observational campaigns, e.g. 2MASS (Jarrett et al. 2003), SLOAN (Stoughton et al. 2002), 
SINGS (Kennicutt et al. 2003) just to mention few.\\
As far as nearby clusters, such as the Virgo cluster and the Coma supercluster, 
multifrequency data for over 3000 galaxies
are collected and distributed via the WEB site "GOLDMine" (Gavazzi et al. 2003).\\
Spectroscopic data are equally invaluable sources of information, but are more difficult to obtain and more 
time consuming. The spectroscopic characterization of the stellar continua provide us with "clocks" on stellar populations, 
while line indices unable us to quantify the chemical evolution of the stars and of the interstellar medium in galaxies. However much
less extensive surveying was carried out in the spectroscopic than in the imaging mode, if one excludes the pioneering work of 
Kennicutt (1992) (K92 hereafter) who first tried to assess the systematic spectral properties of 
nearby galaxies along the Hubble sequence 
and of Jansen et al. (2000) who extended the spectral survey 
of K92 to a large sample of isolated galaxies, later analyzed by Stasinska \& Sodre (2001). 
These spectroscopic surveys were carried out in the 
drift-scan mode, i.e. with the slit sliding over the whole galaxy area.
Spectra taken in this way are representative of the mean galaxies, 
unlike most long slit observations which are dominated by the nuclear light.\\
In 1998, inspired by the work of K92 and of Jansen et al. (2000) we initiated a long term project 
aimed at characterizing spectroscopically the galaxies in the nearest rich cluster: the Virgo cluster. 
In Paper I of this series (Gavazzi et al. 2002a) we analyzed the spectral 
continua based on preliminary 124 spectra obtained until 2001.
Here we present the full set of 333 spectra obtained so far (spring 2003).
They are available in JPG and FITS format at the WEB site GOLDMine (http://goldmine.mib.infn.it) 
(Gavazzi et al. 2003). Their analysis is postponed to Paper III (in preparation).\\
The present paper is organized as follows:
Section 2 describes the surveyed sample, the observational and data reduction techniques.  
Section 3 gives the details of the line measurements, including the deblending of unresolved lines
and the separation of emission from underlying absorption lines. 
The derived spectral parameters are given in Section 4 and 
briefly summarized in Section 5.

\section{Observations}

 
Long-slit spectra of 333 
galaxies, obtained during approximately 50 nights distributed in 6 years (1998-2003) using the $\rm1.93~m$ telescope 
of the Observatoire de Haute Provence (OHP), 
the ESO/3.6 m telescope, the Loiano/1.52 m telescope and the San Pedro Martir (SPM) 2.1 m telescope are presented.
The observations were taken in the "drift-scan" mode: i.e. with the slit, generally parallel
to the galaxy major axis, drifting over the optical surface of the galaxy. \footnote{Galaxies with major axis $>$ 5 arcmin  
were observed with the slit perpendicular to the major axis. 
Few galaxies with both diameters larger than the slit length were observed. However
most of the light from these objects comes from a region corresponding to half the 
($\rm 25^{th}~mag~arcsec^{-2}$) diameters quoted in Tab. \ref{Tab2}, thus well within the slit length.}\\
At ESO we set the guide velocity of the telescope such that during the 
integration time the slit slides one time through the full length of the galaxy. At the remaining observatories 
the drifting was obtained by slewing manually several times the telescope between two extreme positions checked 
on one offset star or on the galaxy itself. 
Not unexpectedly spectra obtained in this way have lower S/N ratio than traditional long-slit spectra
of similar integration time, because a large fraction of the time is spent 
on low surface brightness regions.
Our spectra cover the wavelength range 3600-6800 \AA~ (from [OII] to [SII]) 
with a resolution of $500<R<1000$. 
The spectrograph characteristics are given in Table \ref{Tab1}.
For 6 bright emission line galaxies observed at ESO we used both a low resolution grism and a high resolution red grism. \\
The observations at OHP, Loiano and San Pedro Martir were carried out in approximately 1.5-3 arcsec seeing conditions,
while subarcsecond conditions were often encountered at ESO.
We remark that the present data, owing to the "drift-scan" method 
are marginally affected by the seeing conditions.
The OHP observations were sometimes taken through cirrus, otherwise in transparent or photometric
conditions, while the observations obtained at ESO, SPM and Loiano were transparent or photometric. 
The spectrophotomectric standards Feige 34, Hz 44 and Hiltner 600 (ESO) were observed twice on each night.

\subsection{The sample}

\begin{table*}

\caption{The spectrograph characteristics}
\label{Tab1}
\[
\begin{array}{p{0.15\linewidth}cccccccccc}
\hline
\noalign{\smallskip}
{Telescope} & {Run} & {Spectrograph} & {Disp} & {Disp} & {\Delta\lambda} &{CCD} & {pix} & {Spat. Scale} & {Slit} \\
	    &	  &	       &  {\rm \AA/mm}    &  {\rm \AA/pix} & {\rm \AA}   &    & {\rm \mu m} & \rm "/pix & {\rm arcsec} \\
\noalign{\smallskip}
\hline
\noalign{\smallskip}
OHP/1.93    & 1998  & CARELEC &  260  &  7.0  & 3600-7200 & 512    \times512~TEK      & 27   & 1.17 & 300 \times 2.5		       \\
OHP/1.93    & 1999  & CARELEC &  133  &  1.8  & 3400-7000 & 2048   \times1024~EEV     & 13.5 & 0.58 & 300 \times 2.5		       \\
OHP/1.93    & 2000  & CARELEC &  133  &  1.8  & 3400-7000 & 2048   \times1024~EEV     & 13.5 & 0.58 & 300 \times 2.5		       \\
OHP/1.93    & 2001  & CARELEC &  133  &  1.8  & 3400-7000 & 2048   \times1024~EEV     & 13.5 & 0.58 & 300 \times 2.5		       \\
OHP/1.93    & 2002  & CARELEC &  133  &  1.8  & 3400-7000 & 2048   \times1024~EEV     & 13.5 & 0.58 & 300 \times 2.5		       \\
OHP/1.93    & 2003  & CARELEC &  133  &  1.8  & 3400-7000 & 2048   \times1024~EEV     & 13.5 & 0.58 & 300 \times 2.5		       \\
ESO/3.6(LD) & 2001  & EFOSC2  &  135  &  4.0  & 3380-7520 & 2048  \times2048~LOR      & 15    & 0.16 & 300 \times 1.5		       \\
ESO/3.6(HD) & 2001  & EFOSC2  &  67   &  2.0  & 4700-6770 & 2048  \times2048~LOR      & 15    & 0.16 & 300 \times 1.5		       \\
ESO/3.6(LD) & 2002  & EFOSC2  &  135  &  4.0  & 3380-7520 & 2048  \times2048~LOR      & 15    & 0.16 & 300 \times 1.5		       \\
SPM/2.1     & 2002  & Boller\&Chivens	&  125  &  3.0  & 3900-7000 & 1024  \times1024~SITE3	      & 24   & 0.96 & 300 \times 2.0   \\
LOI/1.52    & 2003  & BFOSC   &  198  &  4.0  & 3600-8900 & 1300   \times1340~EEV     & 20   & 0.58 & 300 \times 2.0		       \\
\noalign{\smallskip}
\hline
\end{array}
\]
\end{table*}

\normalsize
\begin{table*}
\caption{The number of observed spectra}
\label{Tab2}
\[
\begin{array}{p{0.15\linewidth}ccccccccc}
\hline
\noalign{\smallskip}
{Telescope} & {Date} & {Virgo} & {Coma} & {A262} & {Cancer}  &{Centaurus} & {Other} & {Tot} & {Seeing (")}\\
\noalign{\smallskip}
\hline
\noalign{\smallskip}
OHP/1.93    & 5~Mar~1998		 & 9  & -  &  - &  - &  - & - & 9  & 2.0-3.0 \\
OHP/1.93    & 9-15~Mar~1999		 & 22 & -  &  - &  2 &  - & 2 & 26 & 2.0-3.0 \\
OHP/1.93    & 1-6~Feb~2000		 & 27 & -  &  8 & 16 &  - & 2 & 53 & 2.0-3.0 \\
OHP/1.93    & 19-25~Mar~2001		 & 24 & 5  &  - & 1  &  - & - & 30 & 2.0-3.0 \\
OHP/1.93    & 7-13~Mar~2002		 & 51 & 4  &  - & 4  &  - & 1 & 60 & 2.0-3.0 \\
OHP/1.93    & 25~Mar-06~Apr~2003         & 14 & 33 &  - & -  &  - & - & 47 & 2.0-3.0 \\
ESO/3.6(LD) & 23-25~Mar~2001		 & 41 & 1  &  - &  - &  8 & - & 50 & 0.5-2.0 \\
ESO/3.6(HD) & 23-25~Mar~2001		 & (6)& -  &  - &  - &  - & - & (6)& 0.5-2.0 \\
ESO/3.6(LD) & 15-16~Mar~2002		 & 33 & -  &  - &  - &  - & 5 & 38 & 0.5-1.0 \\
SPM/2.1     & 17~Mar~2002		 & 3  & 1  &  - &  - &  - & - & 4  & 1.5-2.0 \\
LOI/1.52    & Jan-Feb~2003	         & 1  & 1  &  - & 14 &  - & - & 16 & 2.0-2.5 \\
\hline
Tot	    &			  & 225+(6)& 45 &  8 & 37 &  8 & 10 & 333+(6)\\
\noalign{\smallskip}
\hline
\end{array}
\]
\end{table*}

\begin{table*}
\caption{Completeness of spectroscopic observations for Virgo cluster members and possible members
regardless of their Hubble type and for late-type galaxies.}
\label{Tab3}
\[
\begin{array}{p{0.15\linewidth}ccccc}
\hline
\noalign{\smallskip}
{$m_{pg}$} & {N~VCC} & {with~z} & {N~Spectra} & {\%} \\
\noalign{\smallskip}
\hline
\noalign{\smallskip}
$\leq$16~all~types	      & 621  & 568 & 223 & (36) \\
$\leq$15~all~types	      & 430  & 427 & 198 & (46) \\
$\leq$14~all~types	      & 252  & 252 & 157 & (62) \\
\noalign{\smallskip}
$\leq$16~late-type    & 323  & 318 & 164 & (51) \\
$\leq$15~late-type    & 244  & 244 & 149 & (61) \\
$\leq$14~late-type    & 151  & 151 & 114 & (75) \\
\hline
\end{array}
\]
\end{table*}

Targets of the present spectrophotometric measurements were primarily selected from the Virgo Cluster Catalog 
(Binggeli et al. 1985: VCC). Among these we observed 225 objects.\footnote{Including 6 additional 
spectra taken with the William Herschel Telescope kindly provided to us by J.M. Vilchez (VCC 324, 334, 562, 841, 848, and 2037) and 
2 spectra taken from the spectral atlas of Kennicutt (1992) (VCC 355 e 1226) 
the total number of Virgo spectra is 233.}
Limiting to the 621 galaxies with $m_p\leq16$ which are 
Virgo cluster members ($V<3000$ or classified as possible members by 
Binggeli et al. 1985, 1993; Gavazzi et al. 1999), 
223 have their spectra measured. At this limiting magnitude the completeness of our spectroscopic 
work is thus 36 \% and it increases to 46 \% at $m_p\leq15$  and to 62 \% at $m_p\leq14$, as listed in 
Table \ref{Tab3}. Most unobserved galaxies are dE and E, therefore the completeness results significantly 
higher among late-type galaxies. In these morphological classes we covered more than 50 \% of 
the galaxies with $m_p\leq16$. At the adopted distance of 17 Mpc (or $\mu=31.1$) $m_p=16$ corresponds to $M_p=-15$, 
thus our survey can be considered as representative of the spectroscopic properties of 
late-type galaxies in the Virgo cluster including dwarf systems.\\
Observations of  CGCG (Zwicky et al. 1961-68) galaxies with $m_p\leq15.7$ in other nearby clusters (45 in Coma+A1367, 37 in Cancer, 8 in the 
A262 and Centaurus clusters and another 10 isolated objects) were taken as fillers when Virgo was not observable. 
These do not form a complete set.\\
Table \ref{Tab2} summarizes the number of obtained spectra in each run and cluster  and the approximate seeing conditions.\\
General parameters derived from the literature for galaxies in the observed sample, along with 
the log-book of the observations, are given in Table 6, arranged as follows:\\
Column 1: Galaxy designation.\\
Column 2, 3: (J2000) celestial coordinates.\\
Column 4:  Heliocentric velocity (from this work or from the literature).\\
Column 5: Cluster membership. The membership to the various sub-units within the Virgo cluster
is according to Gavazzi et al. 1999.\\ 
Column 6: Morphological type (from the VCC or from Gavazzi \& Boselli, 1996).\\
Column 7: S=Seyfert, L=Liner H=HII (from NED).\\
Column 8, 9: Major and minor B band optical diameters (in arcmin). These are consistent with the diameters given in the UGC.\\
Column 10: Distance in Mpc. We assume a distance of 17 Mpc for the members (and possible members) 
of Virgo cluster A, 22 Mpc for Virgo cluster B, 32 Mpc for
objects in the M and W clouds. We adopt a distance of 65 Mpc for A262 and of 33 Mpc for the Centaurus
cluster; 51-74 Mpc for the Cancer cluster, 
according to the membership to the individual sub-groups. Distances of 96 and 91.3 Mpc are assumed 
for Coma and A1367 respectively. We adopt $H_o$ = 75 $\rm km~s^{-1} Mpc^{-1}$.\\
Columns 11, 12, 13:  Total apparent (uncorrected) V, B and H (1.65 $\mu$m) magnitudes. 
These are magnitudes at the $\rm 25^{th} mag~arcsec^{-2}$ isophote obtained consistently with Gavazzi \& Boselli (1996).\\
Column 14: Observing run.\\
Column 15: Photometric quality: P=Photometric; T=transparent; C=thin Cirrus.\\
Column 16: Dispersion.\\
Column 17: Integration time (number of exposures $\times$ individual exposure time).\\

\subsection{Data reduction}

The reduction of the spectra was carried out using standard 
tasks in the IRAF package.\footnote{IRAF is the Image Analysis and
Reduction Facility made available to the astronomical community by the
National Optical Astronomy Observatories, which are operated by AURA,
Inc., under contract with the U.S. National Science Foundation. STSDAS
is distributed by the Space Telescope Science Institute, which is
operated by the Association of Universities for Research in Astronomy
(AURA), Inc., under NASA contract NAS 5--26555.} 
\begin{figure}[!h]
\psfig{figure=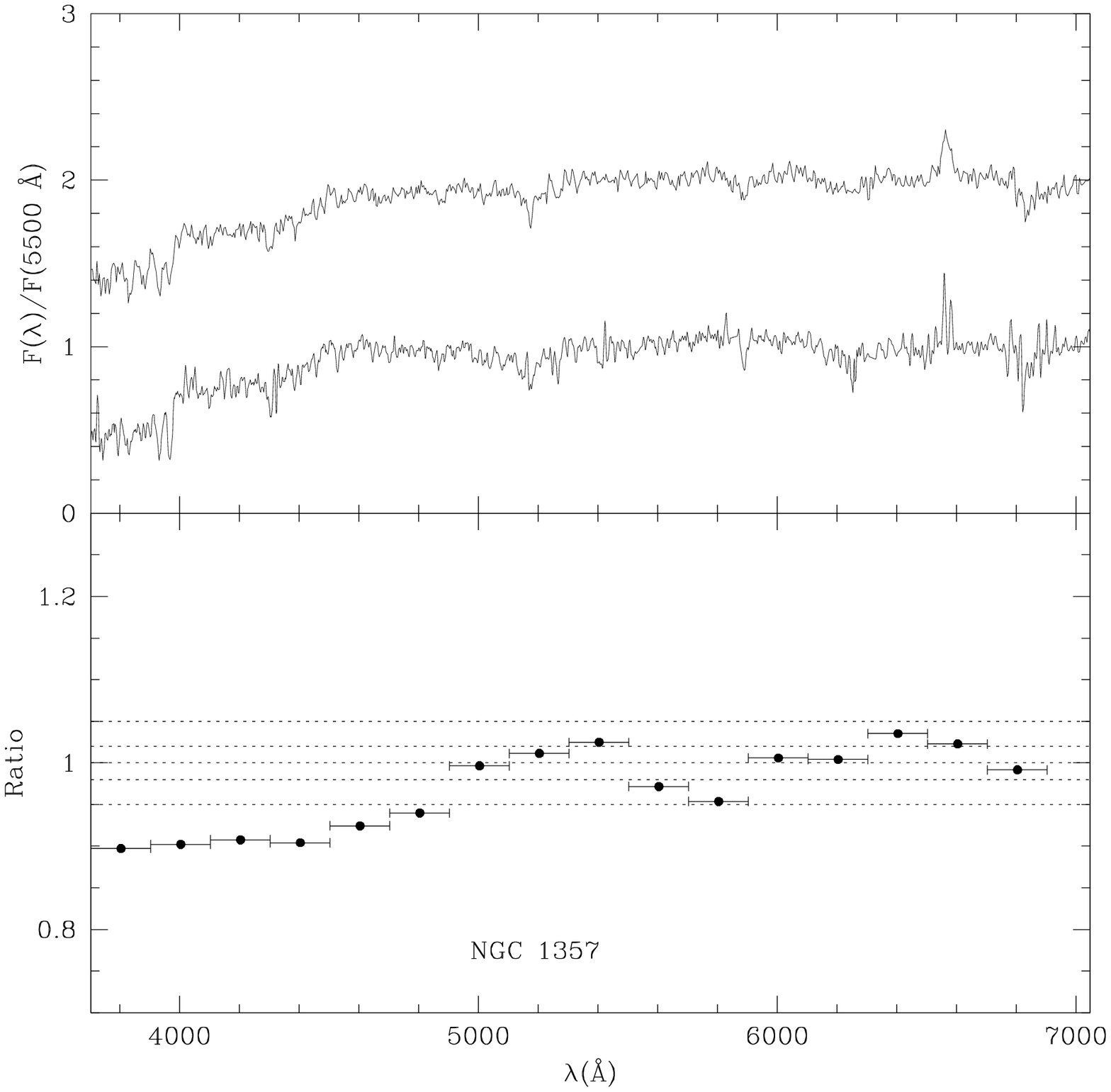,width=9cm,height=9cm}
\psfig{figure=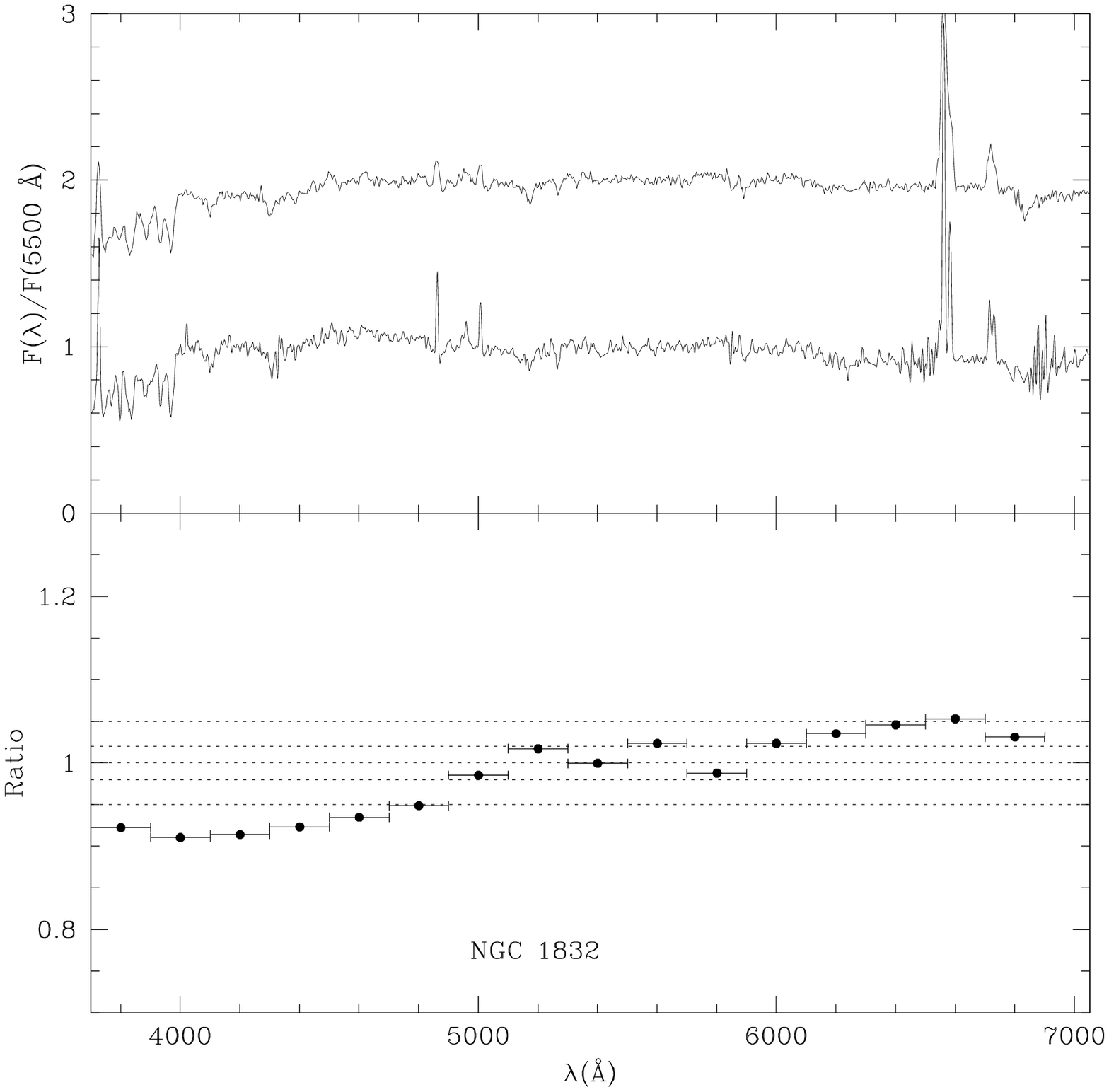,width=9cm,height=9cm}
\caption{Comparison between the spectra of NGC 1357 and 1832 obtained in this work and by K92. 
(top panels). The ratio of the two measurements is given in the bottom panels.}\label{compK}
\end{figure}
\begin{figure}[!h]
\psfig{figure=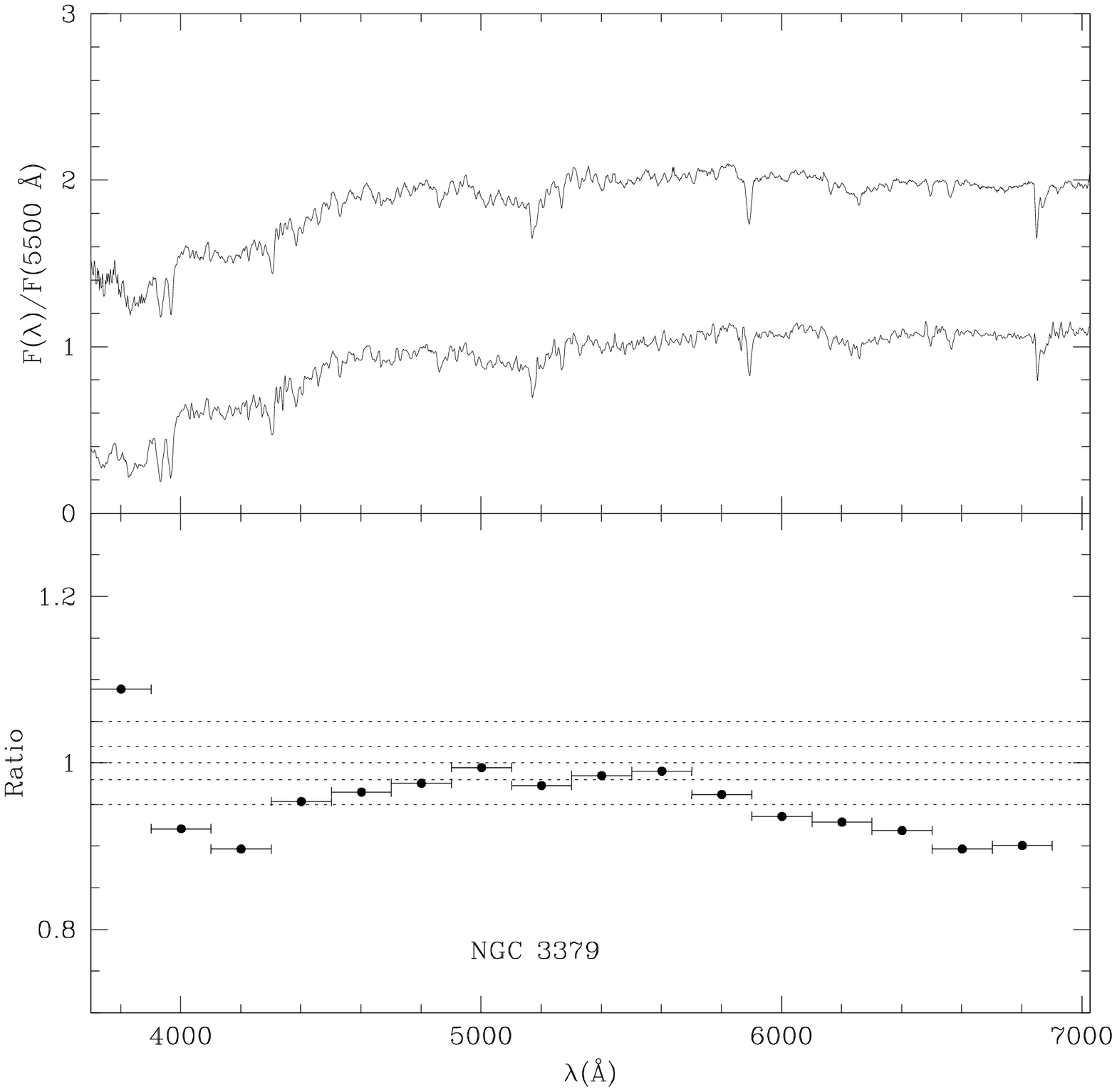,width=9cm,height=9cm}
\psfig{figure=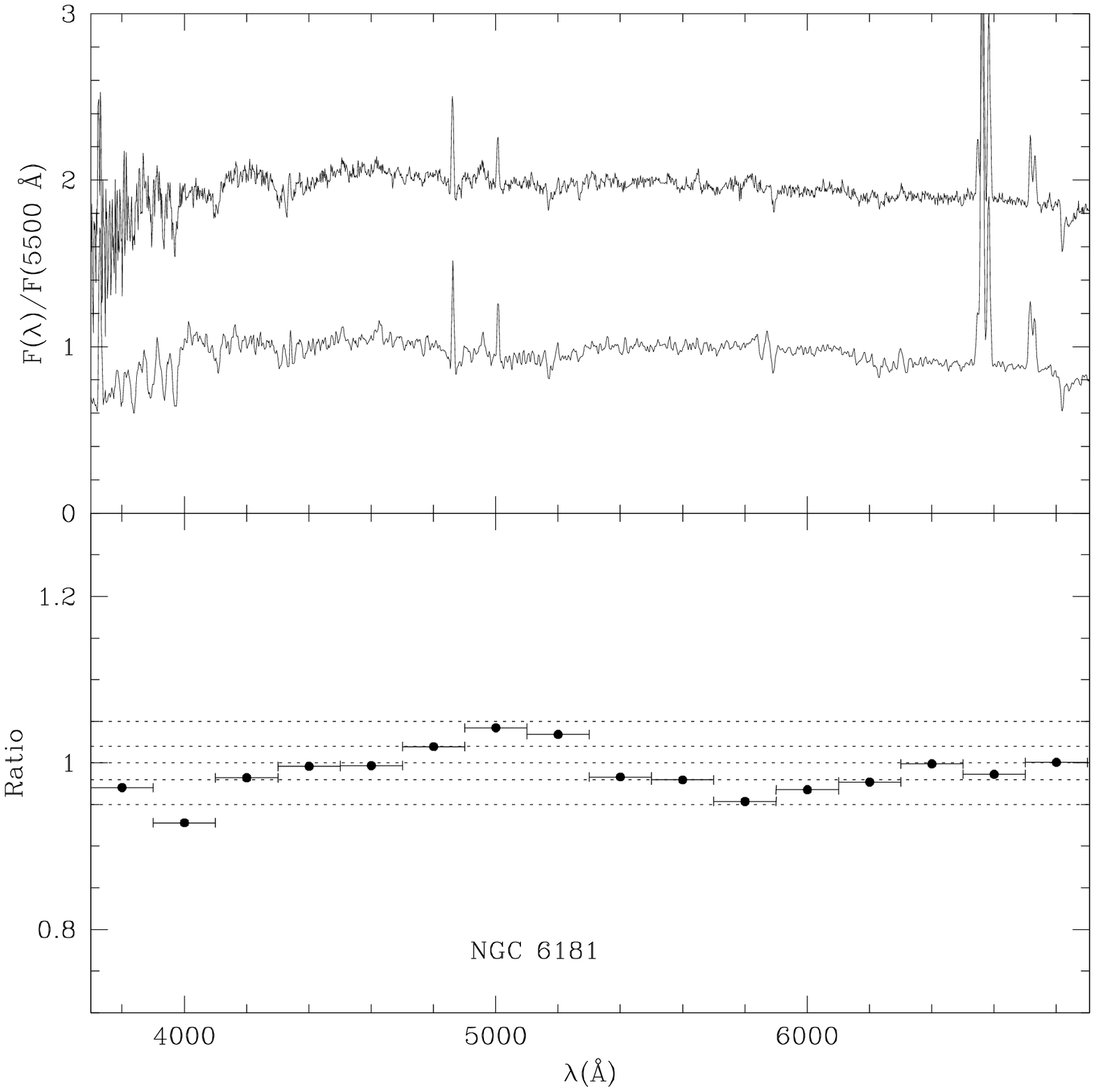,width=9cm,height=9cm}
\caption{Same as Fig.\ref{compK} for NGC 3379 and 6181.}
\label{compK2}
\end{figure}
Visual inspection of the raw images provided us with a list of bad-pixel which were masked from the science frames.
Bias subtraction and flat-field normalization was applied using median of several
bias frames and exposures of quartz lamps. When at least three exposures were obtained for
an object (see ``number of exposures'' in Column 17 of Table 6), they were combined using a median filter, thus removing the cosmic rays. Otherwise
the point-like ones were subtracted using $COSMICRAY$ and the remaining extended features were 
removed under visual inspection of the spectra.\\
The $\lambda$ calibration was carried out using $IDENTIFY-REIDENTIFY-FITCOOR$ on exposures of
He/Ar lamps and the calibration was transferred to the science frames using $TRANSFORM$.
Typical errors on the dispersion solution are of few tenths of $\rm \AA$, as confirmed from the measurements of the sky lines.
The two-dimensional frames were sky subtracted using $BACKGROUND$.  One-dimensional spectra
were obtained integrating the signal along the slit using $APSUM$. 
The apertures were limited to regions where the signal intensity was above
1 $\sigma$ of the sky noise.\\
The flux calibration was achieved using $STANDARD-SENSFUNC-CALIBRATE$ on spectra of  
the standard stars Feige 34, Hz 44 and Hiltner 600 (ESO) taken twice on each night. 
Cubic spline sensitivity functions of 20$^{th}$ order or higher were fit to the calibration spectra, 
allowing the transformation of the
measured intensities into flux densities ($\rm erg~s^{-1} cm^{-2} \AA^{-1}$), including 
the atmospheric extinction correction. However, because during an exposure taken in the ``drift-scan'' mode 
the fraction of light collected by the slit changes with time, an absolute calibration of our science spectra
cannot be achieved. Thus all spectra were normalized to their intensity at $\lambda=5500~$\AA. 
The spectrophotomectric standards were instead used to calibrate the absolute response
of the system as a function of wavelength. The uncertainty of this measurement over the whole
spectral region resulted within 15\%,
as derived comparing the spectra of four galaxies taken in this work with the corresponding 
ones by K92 (see Section \ref{comp_K92}). \\  
\begin{figure}[!t]
\psfig{figure=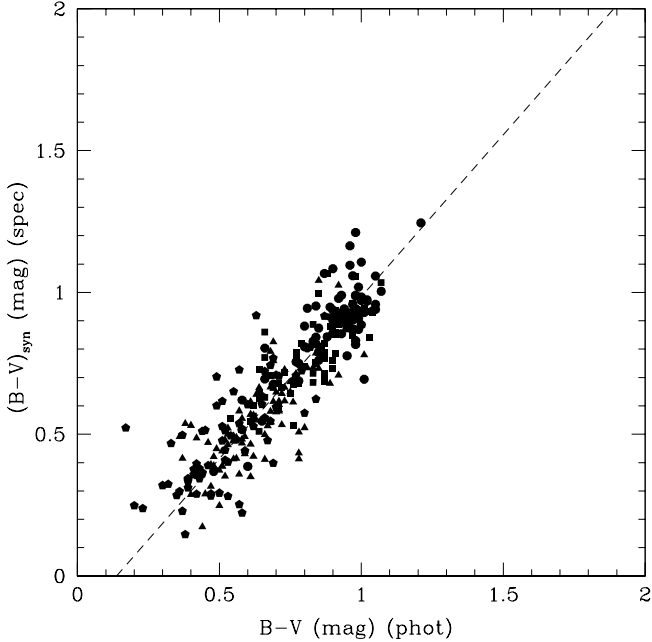,width=9cm,height=9cm}
\caption{Comparison of the synthesized $B-V$ color from our spectra and from photometry.
Unless otherwise specified, in this and following figures dE-E-S0a are represented with
circles; Sa-Sb with squares; Sbc-Scd with triangles; Sd-BCD with pentagons.
The dashed line represents the best linear fit to the data. 
}\label{compBV}
\end{figure}
Three template spectra with high signal-to-noise ratio were selected for being 
representative of absorption-line objects, of weak emission-line objects and of strong emission-line objects
respectively. They were shifted to their rest frame wavelength. 
All the remaining spectra were cross-correlated with one of these template spectra using $FXCOR$, thus providing
their relative redshift. All spectra were shifted to the rest frame wavelength using $DOPCOR$ to better
than 1 $\rm \AA$ and finally they were normalized to their intensity determined in the interval 5400-5600 $\rm \AA$.

\subsection{Comparison with the K92 Atlas}
\label{comp_K92}

Four bright galaxies (NGC 1357, 1832, 3379 and 6181) were measured in common with K92 and their 
comparison is useful to assess the quality of our data.
Figs. \ref{compK} and \ref{compK2} show the spectra of these objects as obtained by us (top spectra) and 
as given by K92 (bottom spectra). The ratio of the two measurements is shown in 
the bottom panel of each figure.  The two sets of data
are found in agreement within 15 \% in the range 3800--6800 $\rm \AA$.

\subsection{Synthetic and photometric color indices}


Synthesized spectroscopic colors $B-V$ and $B-R$ were obtained deconvolving the continua 
with the profiles of the B, V and R Johnson filters.
Since the width of the R filter (5500--7000 $\rm \AA$) exceeds by 200 $\rm \AA$ in the red the 
domain covered by our spectra, $B-R$ is computed deconvolving the Bruzual \& Charlot (1993)
population synthesis models fit to the observed spectra (see Paper I) with the B and R filter profiles.
Fig. \ref{compBV} shows the comparison between the photometric colors $(B-V)_T$
and the $B-V$ color synthesized on the spectra for 312 galaxies.
The two differ by 0.05 mag with an rms scatter of 0.15 mag.\\
The comparison of the synthesized $B-R$ color from our spectra with $(B-V)_T$ color 
from photometry is given in Fig.\ref{B-V_B-R}.  Unless otherwise specified, in this and in the following
figures we plot the best fit linear regression obtained using the $bysector$ method of Feigelson \& Babu (1992).
The results of the linear regression analysis, including the uncertainties in their slope and zero point 
are summarized in Tab. \ref{tabreg}. 

\section{Line measurements}

Under visual inspection to the spectra we carried out a first-order measurement to all lines, 
both in emission and in absorption, using $SPLOT$. This provided a list of fluxes and EWs 
with respect to a user defined continuum level. This preliminary measurement was then refined
as described in the following Sections.
\begin{figure}[!t]
\psfig{figure=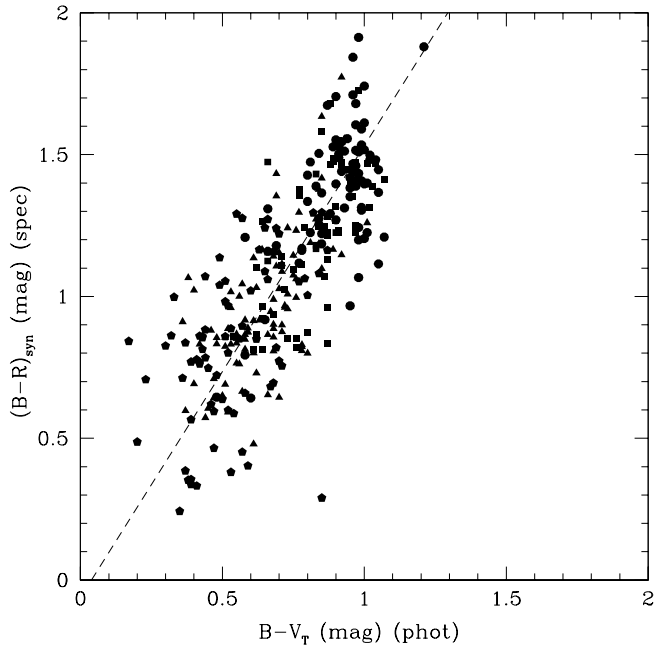,width=9cm,height=9cm}
\caption{Comparison of the synthesized $B-R$ color from our spectra with $(B-V)_T$ 
color from photometry. Same symbols as in Fig.\ref{compBV}.
The dashed line represents the best linear fit to the data. 
}\label{B-V_B-R}
\end{figure}

\begin{table*}
\caption{The bysector linear regression analysis.}
\begin{tabular}{lcc}
\hline
\hline
\noalign{\smallskip}
Regression &R&see Fig.\\
\noalign{\smallskip}
\hline
\noalign{\smallskip}
$(B-V)_{syn}=1.141\pm0.027*(B-V)_T -0.156\pm 0.017$ & 0.86 &\ref{compBV}\\ 
$(B-R)_{syn}=1.593\pm0.052*(B-V)_T -0.062\pm0.033$&0.78&\ref{B-V_B-R}\\
$H_\alpha+[NII1]+[NII2]=0.575\pm0.052*log(L_V/L_\odot) -5.922\pm0.206$ &0.58&\ref{Hacorr}\\
$H_\alpha+[NII1]+[NII2] (sp) =1.189\pm0.054*H_\alpha+[NII1]+[NII2] (ph) -0.322\pm0.057$ &0.75&\ref{compHa}\\
Mg$_2=0.095\pm0.009*log(L_V/L_\odot)-0.701\pm0.043$ & 0.71&\ref{V_Mg2}\\
$NaD=17.031\pm0.975*Mg_2-0.660\pm0.221$ &0.74&\ref{Mg2_NaD}\\
$H_\beta=-8.707\pm1.206*Mg_2+4.045\pm0.218$&-0.54&\ref{Mg2_Hbeta}\\
G$_{4300}=20.365\pm2.406*Mg_2+0.227\pm0.446$&0.60&\ref{Mg2_Gband}\\
$\Delta_{4000}=2.266\pm0.206*Mg_2+0.040\pm0.040$&0.69&\ref{Mg2_delta}\\
\noalign{\smallskip}
\hline
\label{tabreg}
\end{tabular}
\end{table*}

\subsection{Deblending of $H_\alpha$ from [NII]}

\begin{figure}[!t]
\psfig{figure=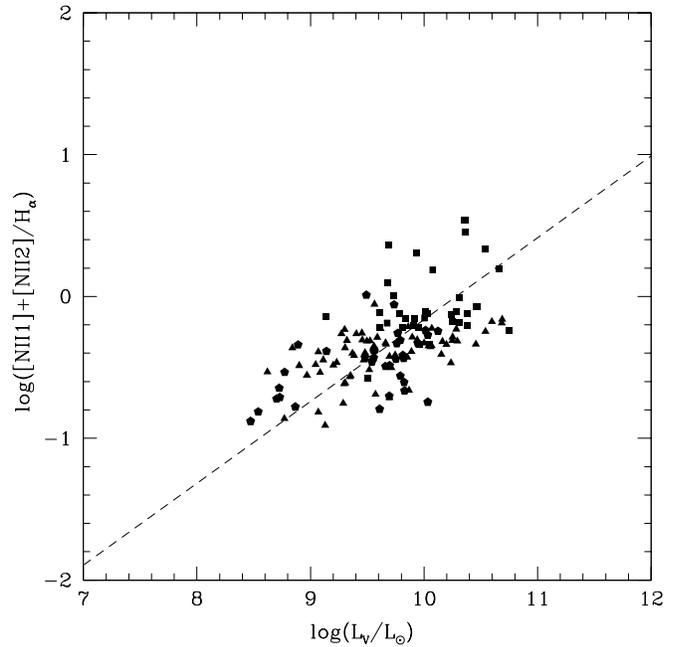,width=9cm,height=9cm}
\caption{$Log(([NII1]+[NII2])/H_\alpha)$ versus V-band luminosity for spectra taken at OHP.
The dashed line represents the best linear fit to the data.}\label{Hacorr}
\end{figure}

$H_\alpha$ ($\lambda 6563$) is bracketed  by the weaker [NII] doublet ([NII1] $\lambda 6548$ and [NII2] $\lambda 6584$). 
The three lines are clearly resolved in the OHP ($R$=1000) spectra, thus for spectra taken at OHP
the measurements of the individual lines is reliable.  
In the lower resolution ($R$=500) spectra taken at ESO, Loiano and SPM the three lines are not well resolved, 
thus the deblending obtained with $SPLOT$ is often inaccurate. 
In most of these cases we could only measure the global 
flux $T$ of the triplet $H_\alpha$+[NII1]+[NII2].
To measure them individually we proceed as follows: 
we calibrate on the OHP spectra the empirical relation between 
(([NII1]+[NII2])/$H_\alpha$) and the V-band luminosity shown in Fig. \ref{Hacorr}, 
which derives from the well known metallicity-luminosity relation (see Raimann et al. 2000). We find:\\
$Log(([NII1]+[NII2])/H_\alpha) = 0.583 (L_V/L_{\odot}) - 5.996$.\\
which, coupled to [NII1] = 0.34 [NII2] provides an estimate of
\begin{figure}[!t]
\psfig{figure=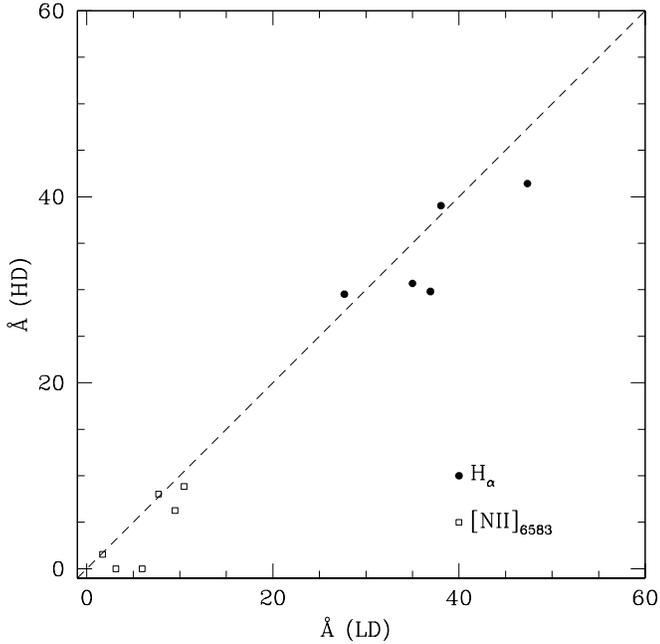,width=9cm,height=9cm}
\caption{Relation between [NII] and H$_{\alpha}$ EW as measured in the High dispersion ESO spectra vs. the same quantities obtained
on the Low dispersion spectra and deblended according to the procedure described in Sect. 3.1. The dashed line represents the one-to-one relation. 
}\label{deblending}
\end{figure}
each individual component [NII1], [NII2] and $H_\alpha$.\\
We apply the above procedure only if [NII2] emission is detected, but not resolved from $H_\alpha$, i.e.
when the line separation is lower than the sum of the individual HWHM.
Otherwise $H_\alpha$ and [NII2] were measured individually and [NII1] was set as 0.34 [NII2].\\ 
The above procedure was checked a posteriori using 6 emission-line galaxies observed at ESO with both the low and 
the high resolution grisms. Fig \ref{deblending} illustrates
the consistency of the EW of [NII2] and $H_\alpha$ obtained on the low resolution spectra 
applying the deblending procedure with those 
directly measured on the HD spectra. The deblended measurements are found overestimated by 20 \% on average.

\subsection{Comparison with $H_\alpha$ from imaging}

\begin{figure}[!h]
\psfig{figure=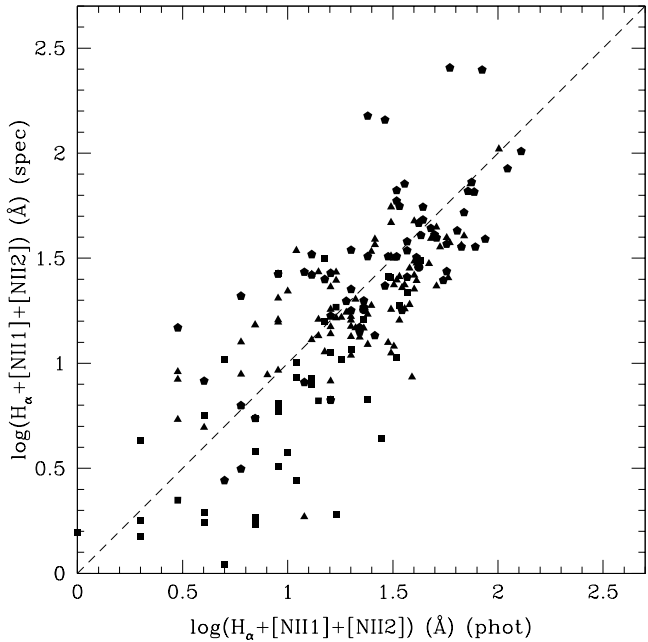,width=9cm,height=9cm}
\caption{Comparison of $H_\alpha$+[NII1]+[NII2] EW derived from imaging and from our spectra. 
Same symbols as in Fig.\ref{compBV}. The dashed line represents the one-to-one relation.}\label{compHa}
\end{figure}

Several (223) galaxies in our sample have their $H_\alpha$+[NII] measured
from imaging (Boselli \&  Gavazzi 2002, Boselli et al. 2002, Gavazzi et al. 2002b).
In spite of the different measuring techniques, 
the equivalent width derived from imaging and from our spectra are found within 0.31 dex $rms$, 
as shown in Fig. \ref{compHa}. This confirms that the ``drift-mode'' spectroscopy
is indeed representative of the entire galaxy, as claimed in the introduction.

\subsection{$[OII]$ $\lambda$3727}

Due to low sensitivity in the blue of CARELEC, only 22\% of the emission line objects 
(EW$H_{\alpha}>$0) observed at OHP have [OII]($\lambda$3727) detected, as opposed to 79\% 
in ESO spectra whose rms noise 
at 4000 \AA~ is half that of OHP spectra.
For the remaining OHP spectra we estimate $3*\sigma$ upper limits to the strength of [OII] 
as $3\times rms_{(3750-4050)}\times 7$, where 7 \AA~ is the mean FWHM of the [OII] lines detected at OHP.

\subsection{Correction for underlying absorption}

\begin{figure}

\psfig{figure=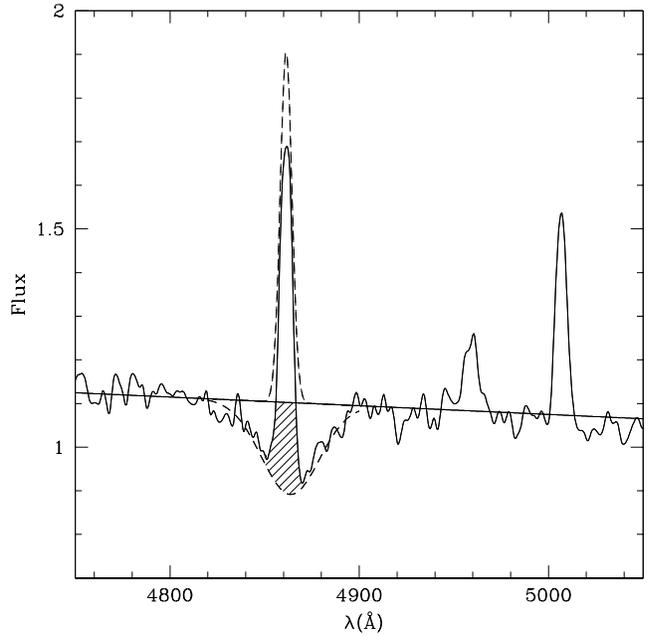,width=9cm,height=9cm}
\caption{Enlargement of the spectrum of VCC 25 to illustrate the deblending of 
$H_\beta$ in emission from the underlying absorption. The observed $H_\beta$ (continuum line) is deblended into a corrected
emission and an absorption component (dotted lines). The shaded region represents the portion of the absorption line
that is added to the emission line to obtain its correction.}\label{under_abs}
\end{figure}
\begin{figure}
\psfig{figure=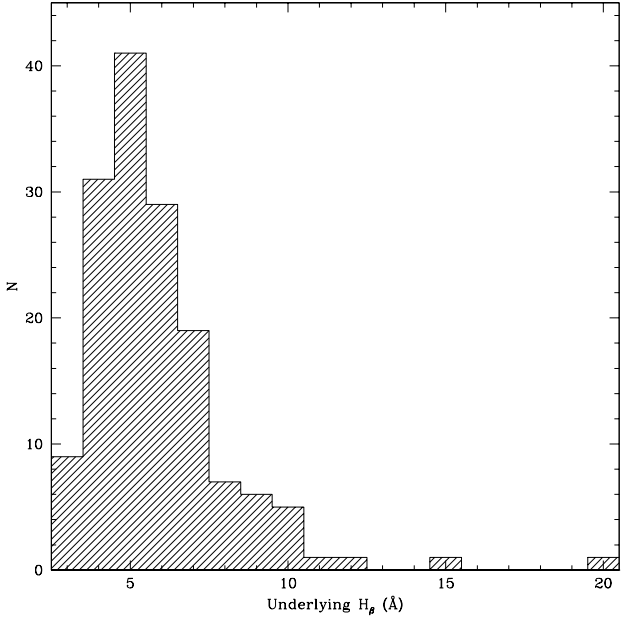,width=9cm,height=9cm}
\caption{Histogram of the underlying EW in absorption at $H_\beta$.}\label{under_histo}
\end{figure}

Most emission line galaxies show evidence for underlying absorption in correspondence to emission lines. 
In particular, out of 174 spectra where we could measure $H\beta$ in emission, in 151 cases
we detect significant $H\beta$ in absorption, and in fewer cases $H\gamma$ and $H\delta$ as well. We
deblended the underlying absorption from the emission lines using a multiple component fitting procedure
written by us in the IRAF environment. 
To do so we measure the emission line and subtract it from the spectra. The resulting 
absorption line is also measured with respect to a reference continuum. 
These two measurements are used as first guess in a
fitting algorithm which fits jointly the emission and absorption lines to the reference continuum (see Fig.\ref{under_abs}).
The value of the emission lines is given in Table 7, that of the underlying absorption in Table 8.\\ 
As shown in Fig. \ref{under_histo} the distribution of the underlying $H_\beta$ is peaked at 5.7 $\pm1.9$ \AA, 
consistently with K92 who reported a mean underlying $H_\beta$ of 5 \AA.\\ 
For objects whose $H_\beta$ was detected in emission but the deblending procedure was not applied (no absorption feature was evident)
a mean additive correction for underlying absorption equal to -1.8 in flux and -1.4 \AA ~in EW was used.
These values correspond to the fraction of the (broader) absorption feature that lies under the emission feature.\\
No mean correction was applied to other lines except $H_\beta$.\\

\subsection{The Balmer decrement ($C_1$)}

\begin{table}[!b]
\caption{Dereddening law relative to $H_\beta$}
\begin{tabular}{ccc}
\noalign{\smallskip}
\hline
\noalign{\smallskip}
Line&$\lambda($\AA$)$&f($\lambda$)-f($H_\beta$)\\
\noalign{\smallskip}
\hline
\noalign{\smallskip}
 $[OII] $& 3727 & 0.31 \\ 
 $H_\delta$ & 4101 & 0.20 \\ 
 $H_\gamma$ & 4340 & 0.13 \\ 
 $H_\beta$ & 4861 & 0 \\ 
 $[OIII]$ & 4958 & -0.02 \\ 
 $[OIII]$ & 5007 & -0.03 \\ 
 $[NII]$ & 6548 & -0.33 \\ 
 $H_\alpha$ & 6563 & -0.33 \\ 
 $[NII]$ & 6584 & -0.34 \\ 
 $[SII]$ & 6717 & -0.37 \\ 
 $[SII]$ & 6731 & -0.37 \\ 
 \noalign{\smallskip}
\hline
\noalign{\smallskip}
\label{lequeux}
\end{tabular}
\end{table}

\begin{figure}
\psfig{figure=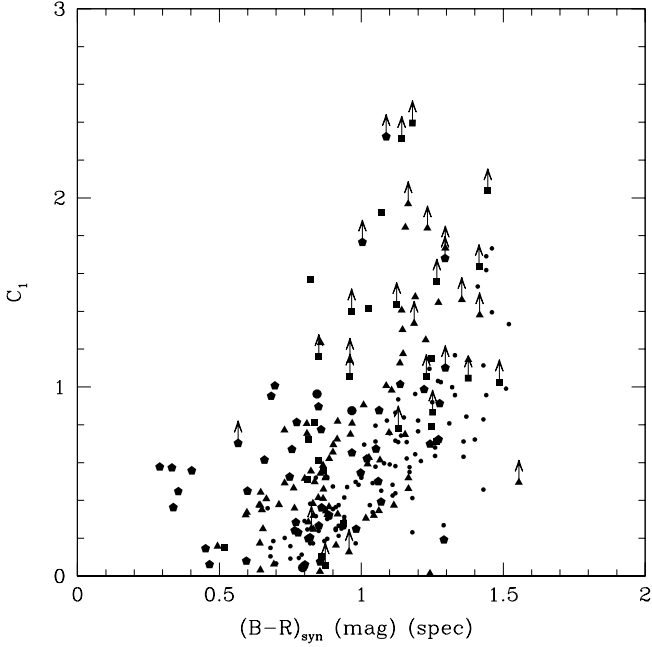,width=9cm,height=9cm}
\caption{$C_1$ versus $(B-R)_{syn}$ (excluding Seyfert objects).
Same symbols as in Fig.\ref{compBV} with the addition of 
galaxies observed by Jansen et al. (2000) (small dots).
}\label{C1_B-R}
\end{figure}

From $H_\beta$ corrected for underlying absorption (Sect 3.4) and $H_\alpha$ corrected for 
deblending from [NII] (Sect 3.1) we evaluate the Balmer decrement:\\
{\small
$C_1=(log(H_\alpha/H_\beta)_{theor}-log(H_\alpha/H_\beta)_{obs})/(f(H_\alpha)-f(H_\beta))$}\\
(in the current notation: A($H_\beta$)=2.5*$C_1$).\\
The ratio $log(H_\alpha/H_\beta)_{theor}$ depends on the electron density and on the gas temperature. 
Assuming T=10000K and n=100 $\rm e/cm^3$, as in Osterbrock (1989) case B,  $(H_\alpha/H_\beta)_{theor}=2.86$ holds.\\ 
The corrected line fluxes are derived, relative to $H_\beta$, using $C_1$  and the  reddening 
function $f(\lambda)$ of Lequeux et al. (1979) (see Table \ref{lequeux}) based on the extinction law of Whitford (1958)
\footnote{ We have compared the $f(\lambda)$ of Lequeux et al. (1979) with
the derivation of Cardelli et al. (1989), which assumes the extinction law of Seaton (1979) and
of Savage \& Mathis (1979). The two functions are in agreement (in the optical range of our interest):
differences are $\leq$ 0.03 from [OII] to H$\alpha$ (included); 0.05 at [SII].}.\\
When EW$H_\alpha$ $>$ 1 \AA~ but $H_\beta$ is undetected we derive a $3*\sigma$
lower limit to $C_1$ using (Buat et al. 2002):\\
$H_\beta<3\times rms_{(4500-4800)}\times H_\alpha HWHM$\\
assuming that $H_\alpha$ and $H_\beta$ have similar HWHM (Half Width Half Maximum).\\
Fig.\ref{C1_B-R} shows the obtained $C_1$ as a function of the synthetic $B-R$ color index 
for our objects (coded according to the morphological type) and for 
galaxies observed by Jansen et al. (2000) and analyzed by Stasinska \& Sodre (2001). 
Our data confirm the positive correlation between the two quantities: i.e. increasing 
$C_1$ with increasing $B-R$. However the dispersion appears higher and the relation steeper than in 
Stasinska \&  Sodre (2001). Notice that 
these authors did not measure the underlying absorption at $H_\beta$.

\section{Results}

The 333 (rest-framed and normalized) spectra obtained in this work are illustrated in Fig. \ref{spec}. \\
The ESO spectra
are given in the range 3600--6800 $\rm \AA$. 
The OHP spectra, noisier 
in the blue, were resampled with a step of 5 $\rm \AA$ and are given from 3850 to 6800 $\rm \AA$, 
unless the strength of the [OII] line was higher than 3 times
the noise determined locally near the line (see Section 3.3). 
The flux scale of Fig. \ref{spec} is given in three intervals: 0.2 -- 2; 0.2 -- 5, 0.2 -- 15 according
to the intensity of the brightest lines. 
The additional 6 high dispersion spectra obtained at ESO are given in the last page of Fig. \ref{spec}.\\
All spectra presented in Fig. \ref{spec} are available in the FITS format at the WEB site GOLDMine 
(http://goldmine.mib.infn.it/).\\

\subsection{Emission lines}

The (corrected) emission lines parameters are listed in Table 7:\\
Column 1: Galaxy identification.\\
Column 2: Balmer decrement $C_1$ (or lower limit).\\
Column 3-13: Line intensities corrected for Balmer decrement, normalized to $H_\alpha$.\\
Column 3: [OII] ($\lambda 3727)$ (or upper limit).\\
Column 4: $H_\delta $ ($\lambda 4101$).\\
Column 5: $H_\gamma $ ($\lambda 4340$).\\
column 6: $H_\beta$ ($\lambda 4861$).\\
Column 7: [OIII] ($\lambda 4959$).\\
Column 8: [OIII] ($\lambda 5007$).\\
Column 9: [NII] ($\lambda 6548$).\\
Column 10: $H_\alpha$ ($\lambda 6563$).\\
Column 11: [NII] ($\lambda 6584$).\\
Column 12: [SII] ($\lambda 6717$).\\
Column 13: [SII] ($\lambda 6731$).\\
Column 14-24: Equivalent widths ($\rm \AA$).\\
Column 14: [OII] ($\lambda 3727$) (or upper limit).\\
Column 15: $H_\delta $ ($\lambda 4101$).\\
Column 16: $H_\gamma$ ($\lambda 4340$).\\
Column 17: $H_\beta$ ($\lambda 4861$).\\
Column 18: [OIII] ($\lambda 4959$).\\
Column 19: [OIII] ($\lambda 5007$).\\
Column 20: [NII] ($\lambda 6548$).\\
Column 21: $H_\alpha$  ($\lambda 6563$).\\
Column 22: [NII] ($\lambda 6584$).\\
Column 23: [SII] ($\lambda 6717$).\\
Column 24: [SII] ($\lambda 6731$).\\
Column 25: Notes.\\

\subsection{Balmer absorption lines}

\begin{figure}
\psfig{figure=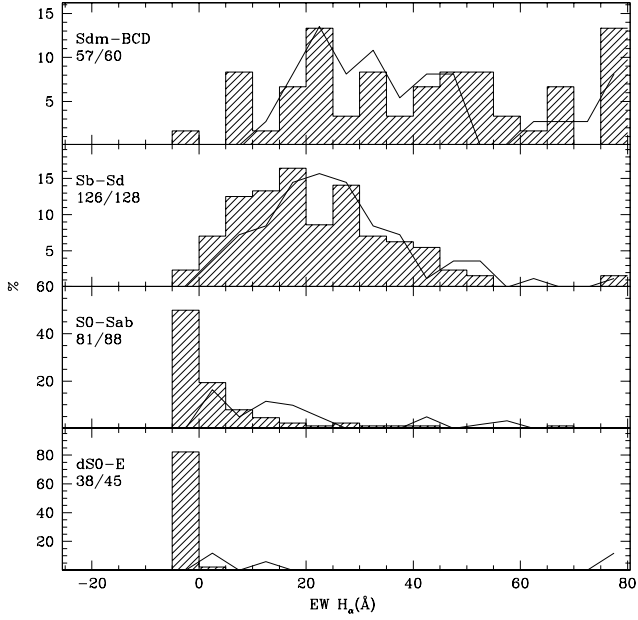,width=9cm,height=9cm}
\caption{Distribution of H$_{\alpha}$ E.W. in four intervals of Hubble type.  Positive E.W. 
represent emission lines, negative values represent absorption lines.
The continuum line represents the frequency distribution obtained with the 
Jansen et al. (2000) data.}\label{Haew_dist}
\end{figure}

\begin{figure}
\psfig{figure=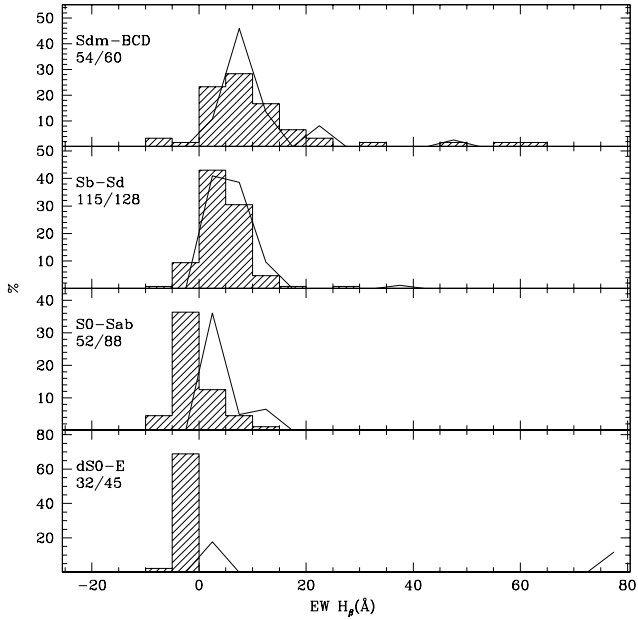,width=9cm,height=9cm}
\caption{Same as in Fig.\ref{Haew_dist} for H$_{\beta}$ E.W.}\label{Hbew_dist}
\end{figure}

\begin{figure}
\psfig{figure=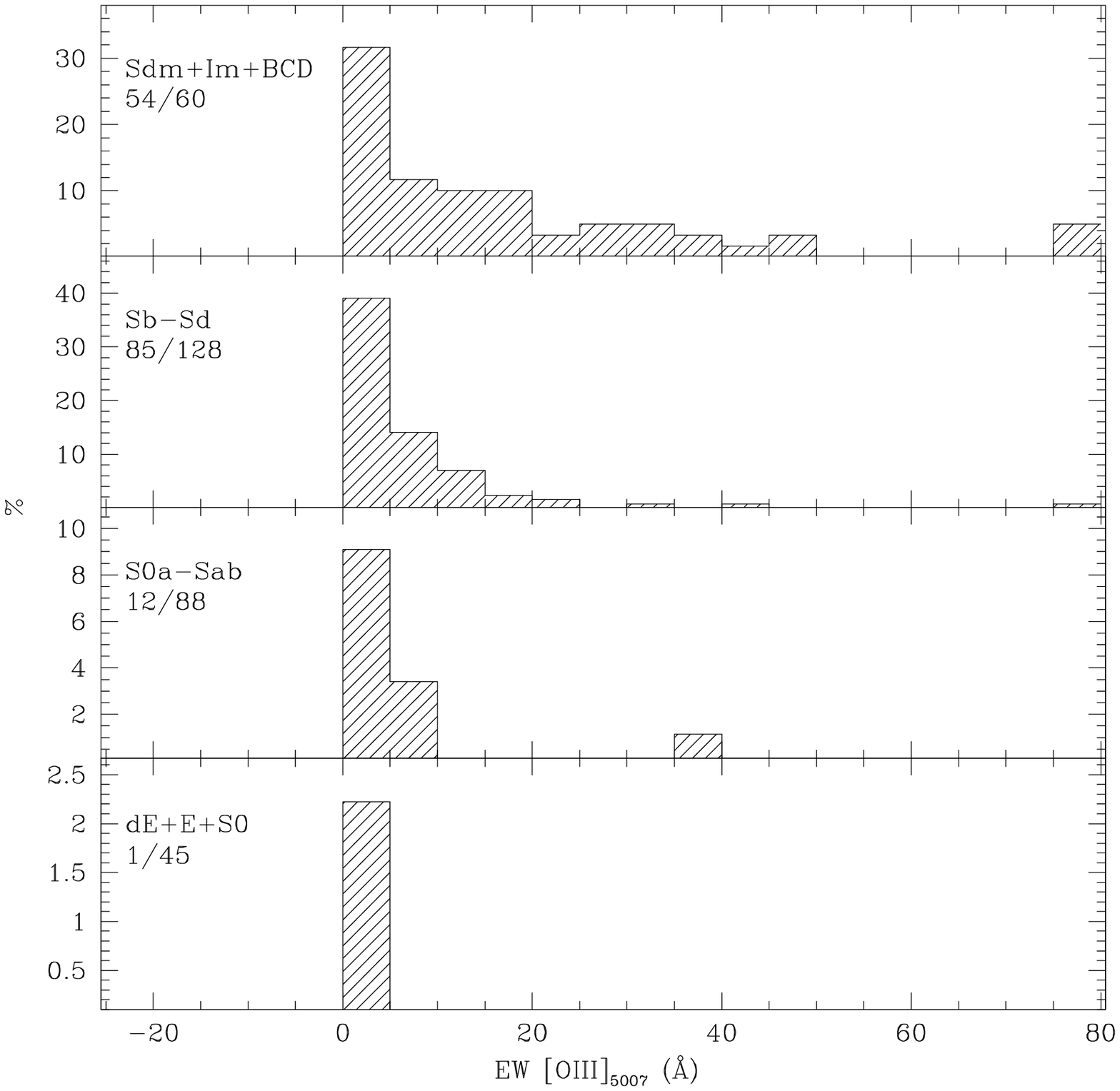,width=9cm,height=9cm}
\caption{Same as in Fig.\ref{Haew_dist} for [OIII2] E.W.}\label{OIII2ew_dist}
\end{figure}

Whenever an absorption feature is detected in correspondence of 
Balmer lines (either alone or deblended from emission, as discussed in Sect. 3.4)
its EW is listed in Table 8 as follows:\\
Column 1: Galaxy identification.\\
Column 2: $H_\delta $ ($\lambda 4101$).\\
Column 3: $H_\gamma $ ($\lambda 4340$).\\
Column 4: $H_\beta$ ($\lambda 4861$).\\
Column 5: $H_\alpha$  ($\lambda 6563$).\\
Column 6: Notes.\\ 
Frequency distributions of the EWs of the principal (emission and absorption) lines 
($H_\alpha$, $H_\beta$ and [OIII] ($\lambda 5007$)) in 4 intervals of Hubble type are given in 
Figs.\ref{Haew_dist}, \ref{Hbew_dist} and \ref{OIII2ew_dist} respectively. 
For each Hubble type interval, the total number of objects and the number of objects with a given 
measured line (either in emission or in absorption) is labeled in each panel. 
The continuum lines represent similar frequency distributions obtained with the 
Jansen et al. (2000) data. 
If one excludes the absorption lines (negative E.W.) that Jansen et al. (2000) did not measure,  
whereas they are included in our analysis, the two distributions appear consistent one another. In fact
the probability that the two distributions are derived from the same parent populations is:
$>$ 62 \% for S0-Sab, $>$ 97 \% for Sb-Sd and $>$ 57 \% for Sdm-BCD, as derived from the Kolmogorov-Smirnov test
(dS0-E are excluded from this analysis due to the poor statistics).
The implications of this finding is that, to the first order, galaxies in rich clusters do not have
emission line properties dramatically different from isolated galaxies. Small differences, if any, 
require a more subtle analysis to be identified, which is postponed to Paper III.

\subsection{Absorption line indices}

The absorption lines indices are derived for early-type objects (dE-dS0-E-S0) and are listed in Table 9. 
They are derived according to the Lick system (Worthey et al. 1994) as the result of
the integration in defined bands. Continua are determined
on both sides of the line under measurement, and averaged. The line strength (or EW) 
is given as the difference between the integral in the interval containing the line 
and in the adjacent continuum.\footnote{The integration has been carried in the Lick system, 
but the indices were not corrected using Lick standards. Notice that 
the EW$H\beta$ listed in Table 9
do not correspond with those in Table 8 for galaxies in common between the two Tables 
(e.g. elliptical galaxies with $H\beta$ in 
absorption) because they are obtained with different measuring techniques.}\\
Table 9 is organized as follows:\\
Column 1: Galaxy identification.\\
Column 2: Calcium Break ($\Delta_{4000}$) $(\lambda 4000)$ in $\rm mag$.\\
Column 3: $G_{4300}$ $(\lambda \lambda ~4283-4317)$ in \AA.\\
Column 4: $H\beta$ $(\lambda \lambda ~4849-4877)$ in \AA.\\
Column 5: $Mg_2$ $(\lambda \lambda ~5156-5197)$ in $\rm mag$.\\
Column 6: $NaD$ $(\lambda \lambda ~5879-5911)$ in \AA.\\
\begin{figure}[!t]
\psfig{figure=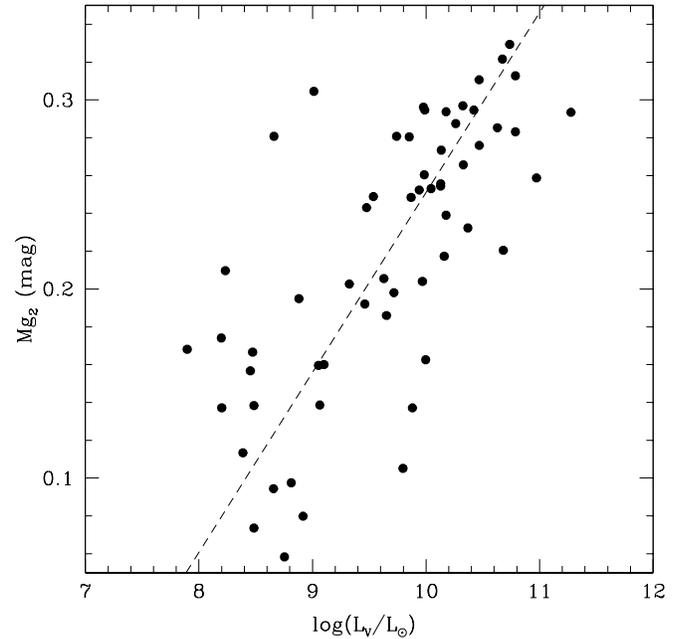,width=9cm,height=9cm}
\caption{The $Mg_2$ absorption index vs. V band luminosity for dE-dS0-E-S0. 
The dashed line represents the best linear fit to the data.}\label{V_Mg2}
\end{figure}
\begin{figure}[!t]
\psfig{figure=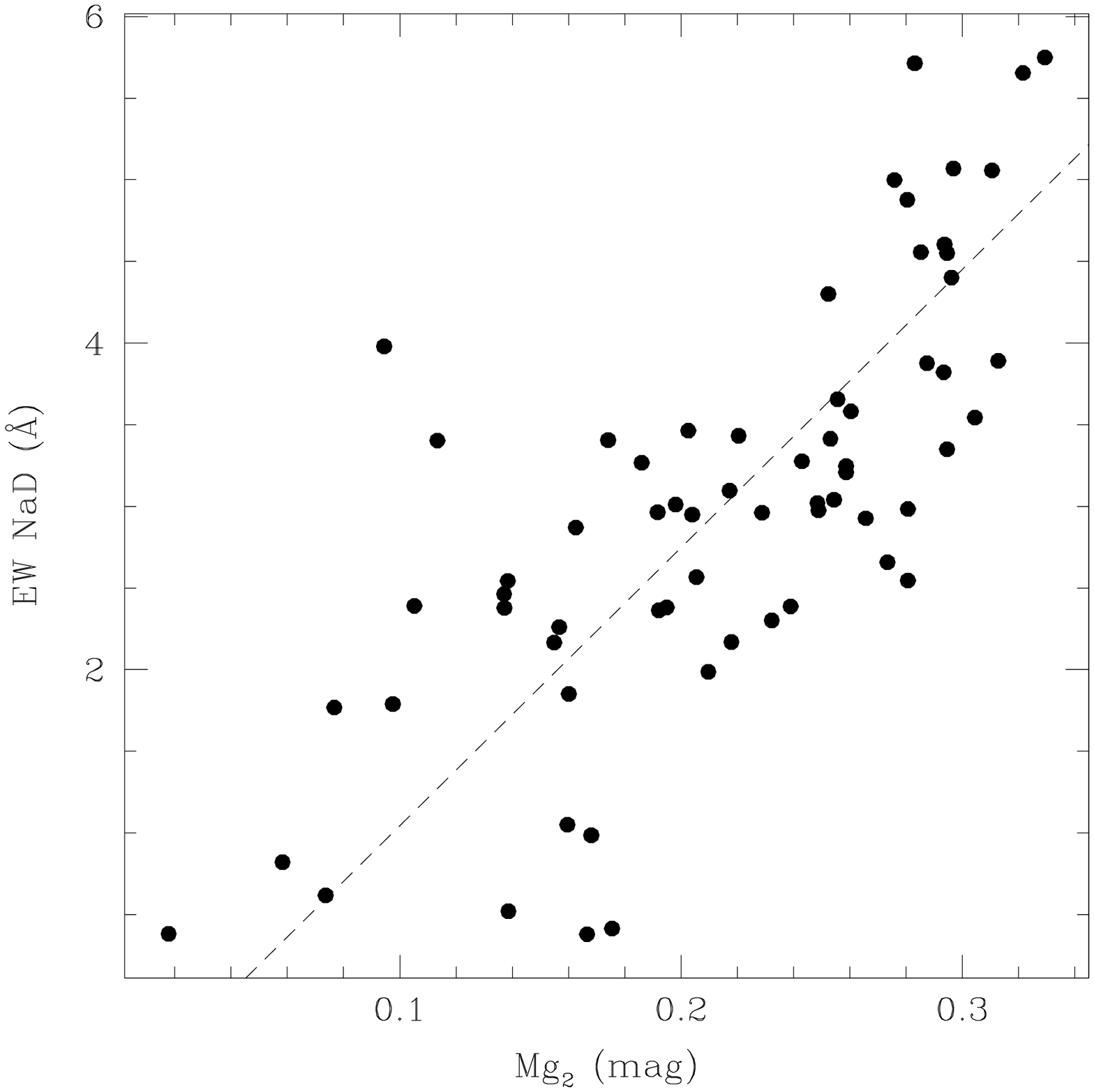,width=9cm,height=9cm}
\caption{The $Mg_2$ absorption index vs. the NaD absorption index  for dE-dS0-E-S0. 
The dashed line represents the best linear fit to the data.}\label{Mg2_NaD}
\end{figure}
\begin{figure}[!h]
\psfig{figure=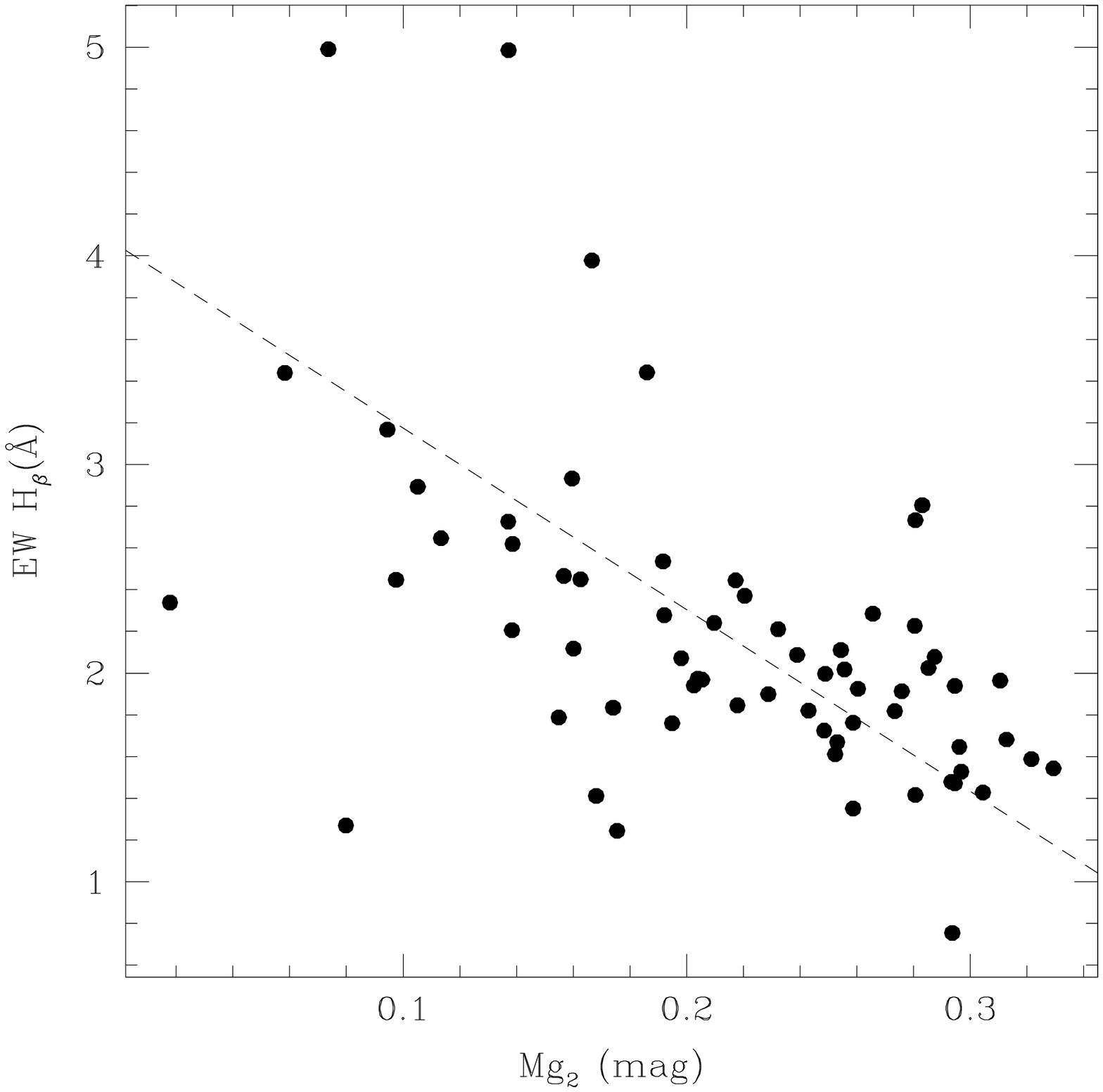,width=9cm,height=9cm}
\caption{The $Mg_2$ absorption index vs. the H$_{\beta}$ absorption index  for dE-dS0-E-S0. 
The dashed line represents the best linear fit to the data.}\label{Mg2_Hbeta}
\end{figure}
\begin{figure}[!h]
\psfig{figure=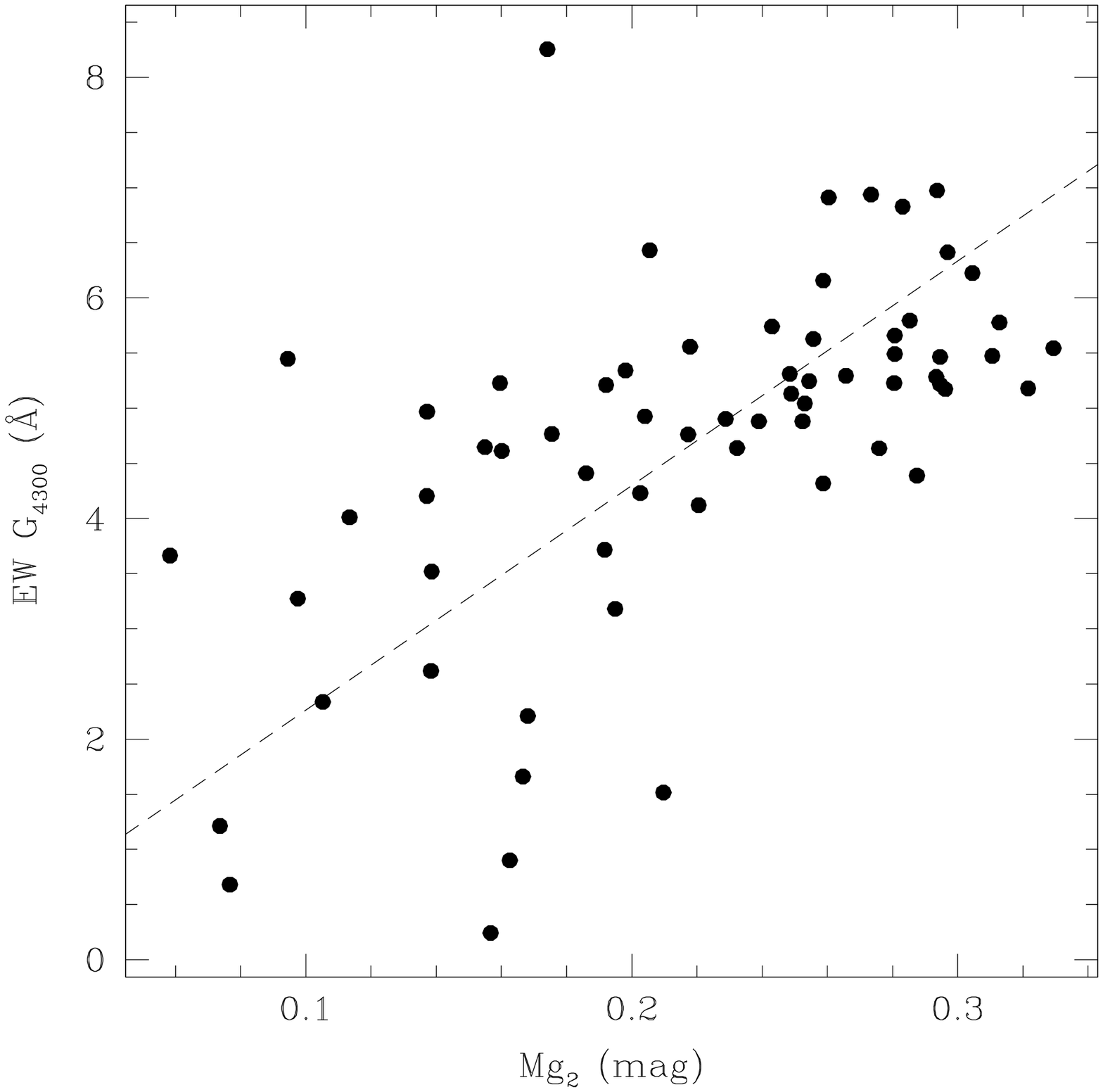,width=9cm,height=9cm}
\caption{The $Mg_2$ absorption index  vs. the $G_{4300}$ absorption index for dE-dS0-E-S0. The dashed 
line represents the best linear fit to the data.}\label{Mg2_Gband}
\end{figure}
\begin{figure}[!h]
\psfig{figure=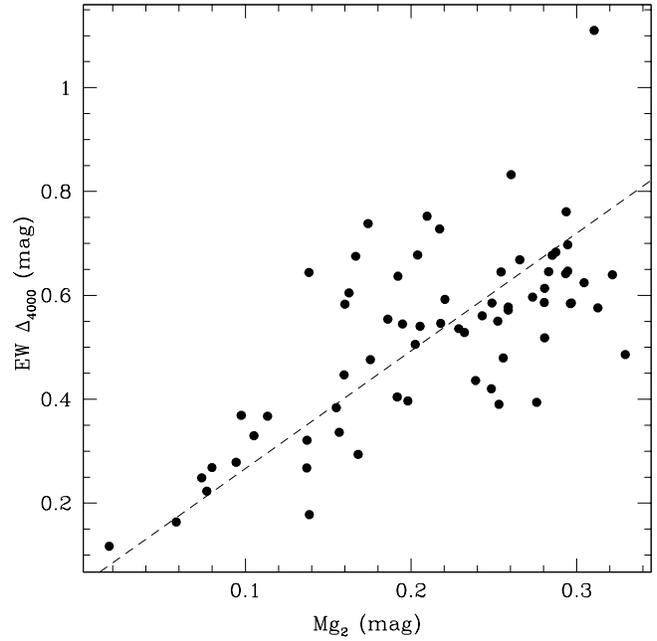,width=9cm,height=9cm}
\caption{The $Mg_2$ absorption index  vs. the Calcium break ($\Delta_{4000}$)  for dE-dS0-E-S0. The dashed 
line represents the best linear fit to the data.}\label{Mg2_delta}
\end{figure}
The principal metallicity index, $Mg_2$ is plotted in Fig.\ref{V_Mg2}
as a function of the V band luminosity, confirming the well known increase of metallicity with 
luminosity (Bica and Alloin 1987). Other absorption line indices, namely
NaD, $G_{4300}$ and $\Delta_{4000}$ are plotted in Figs.\ref{Mg2_NaD}, \ref{Mg2_Gband}
and \ref{Mg2_delta} as a function of $Mg_2$, showing the expected correlation with metallicity. 
On the opposite the strength of $H\beta$ in absorption shows the reverse trend with 
$Mg_2$ (Fig.\ref{Mg2_Hbeta}), as this line is more sensitive to the age than to the metallicity
of stellar populations (Worthey et al. 1994). 
The significance of these correlations can be estimated from
the uncertainty in the slope and zero point listed in Tab. \ref{tabreg}. 
Beside the mentioned dependence on luminosity, 
no significant differences are found among dE and dS0 or among E and S0.

\section{Summary and conclusions}

Using 5 middle-size telescopes for 50 nights distributed in 6 years we obtained drift-scan spectra (3600-6800 \AA) 
with $500<R<1000$ for 333 galaxies in nearby clusters. The majority (225) where secured for galaxies 
in the Virgo cluster.
The observations can be considered representative of the spectral properties of giant and dwarf 
galaxies in this cluster, 
as the completeness achieved at $M_p=-15$ is 36 \% for all types and 51\% for late-type galaxies.\\
Here we present the individual spectra reduced to their rest-frame wavelength and normalized to 
their intensity at 5500 \AA.\\
Intensities (corrected for dereddening) and EWs are derived for the principal lines both in emission 
and in absorption. 
Special care is devoted to deblending of $H_\alpha$ from the [NII] doublet and of emission 
lines from underlying absorption. 
In the case of $H\beta$ we measured underlying absorption 
87 \% of the times we detected emission, with a mean EW of 5.7 \AA.\\
For early-type galaxies we derive the Lick absorption line indices.\\
The complete line analysis is postponed to a forthcoming Paper III where the metallicity indicators 
will be derived and analyzed.\\
The comparison of the line properties of our cluster sample with those of 200 isolated galaxies 
observed by Jansen et al (2000) 
with a similar experimental setup will allow to study the influence of the cluster environment 
on the interstellar medium and on 
the stellar population of galaxies. This analysis will help shading light on the elusive 
processes that made cluster galaxies 
evolve separately from their isolated counterparts.   

\acknowledgements

This research has made use of the 
NASA/IPAC Extragalactic Database (NED) and of the GOLDMine database. NED is operated 
by the Jet Propulsion Laboratory, California Institute of Technology, under contract with the
National Aeronautics and Space Administration. GOLDMine is operated by the Universita' degli Studi di Milano-Bicocca.

\newpage

\begin{figure*}
\psfig{figure=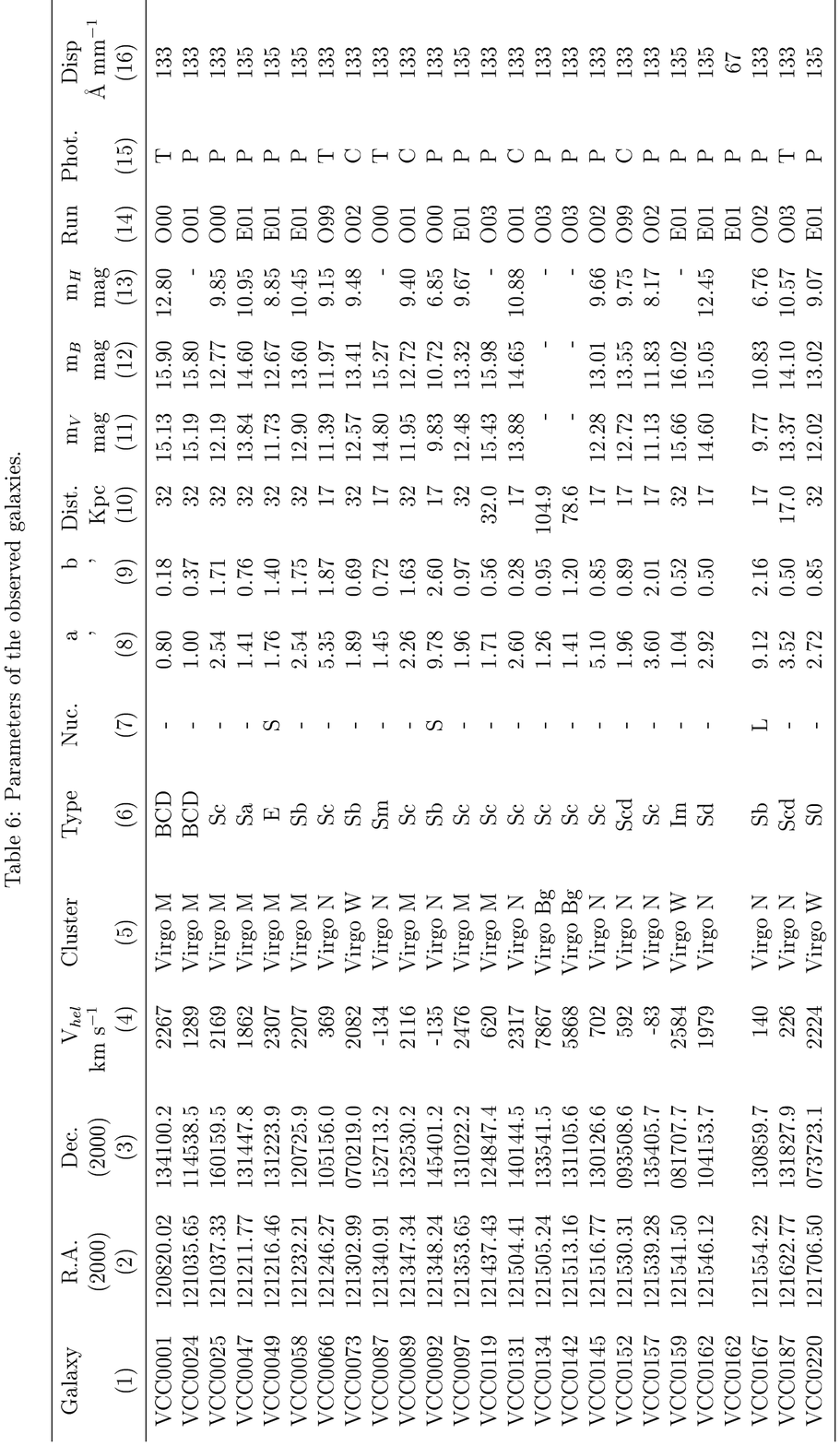,width=16cm,height=23cm}
This is one page sample. The entire table is avaible at 
URL http://goldmine.mib.infn.it/papers/vccspec\_2.html
\end{figure*}
\setcounter{table}{5}
\begin{figure*}
\psfig{figure=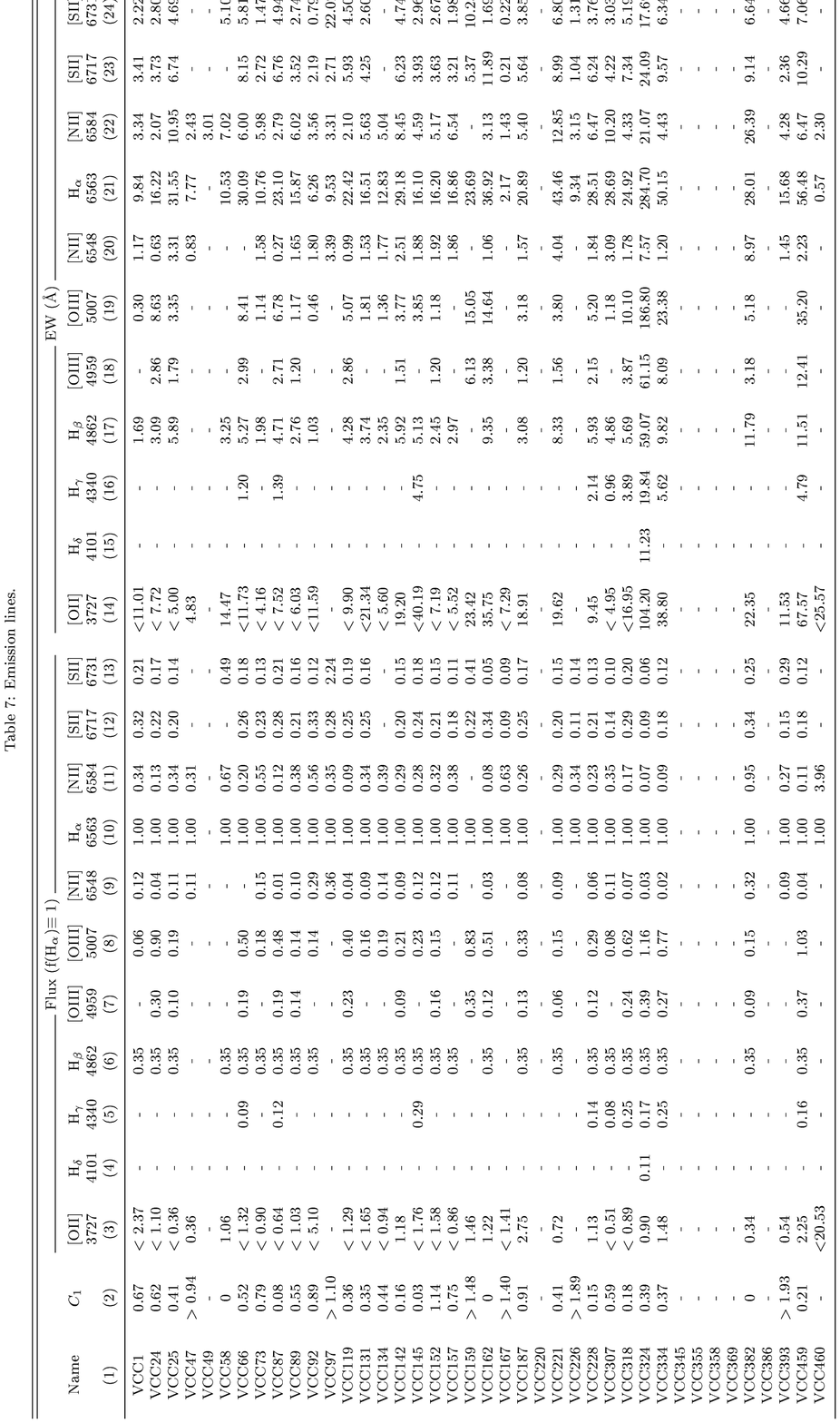,width=16cm,height=23cm}
This is one page sample. The entire table is avaible at 
URL http://goldmine.mib.infn.it/papers/vccspec\_2.html
\label{lines_tab}
\end{figure*}

\begin{figure*}
\psfig{figure=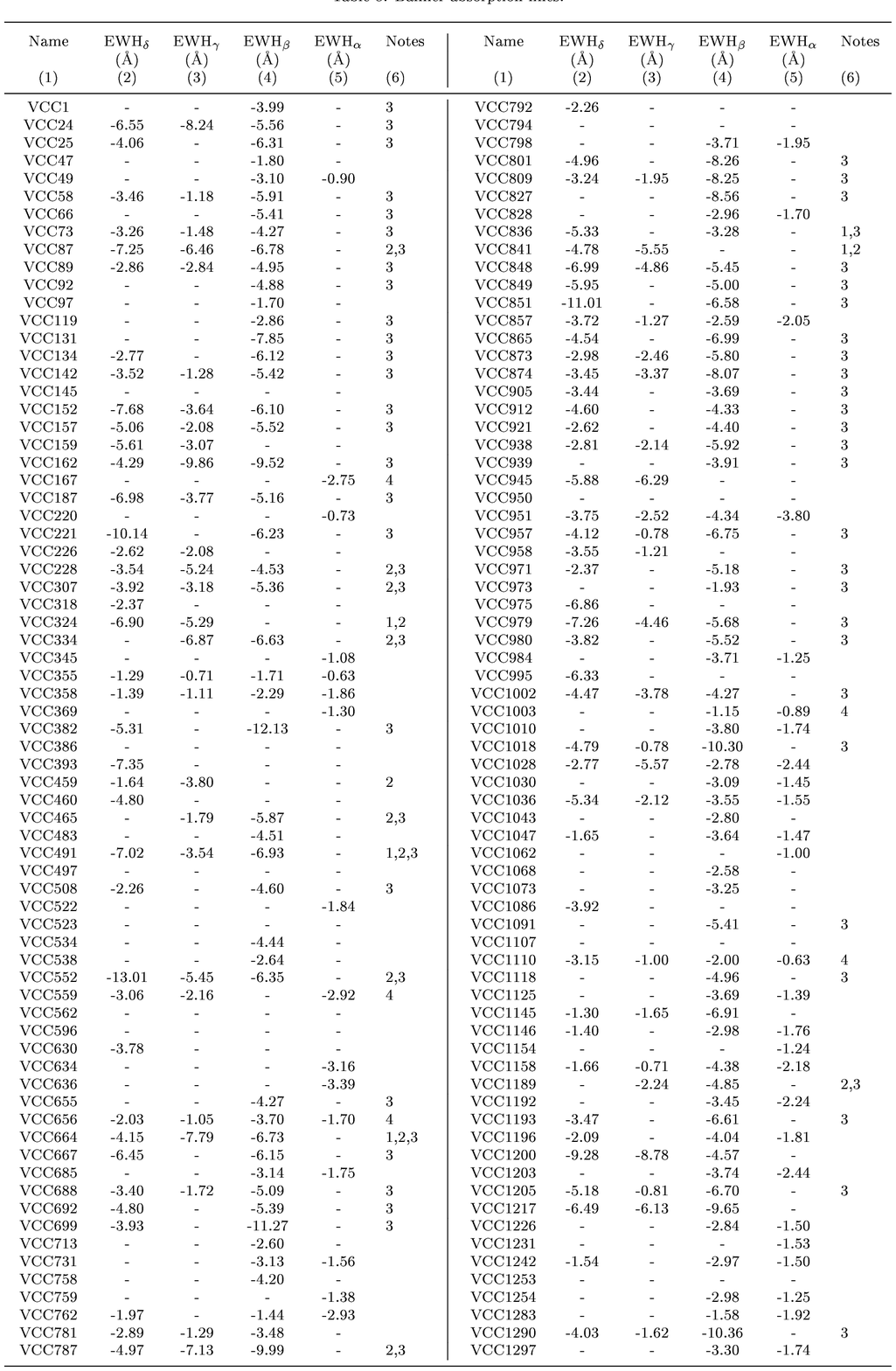,width=16cm,height=23cm}
This is one page sample. The entire table is avaible at 
URL http://goldmine.mib.infn.it/papers/vccspec\_2.html
\label{Bal_Abs}
\end{figure*}

\begin{figure*}
\psfig{figure= 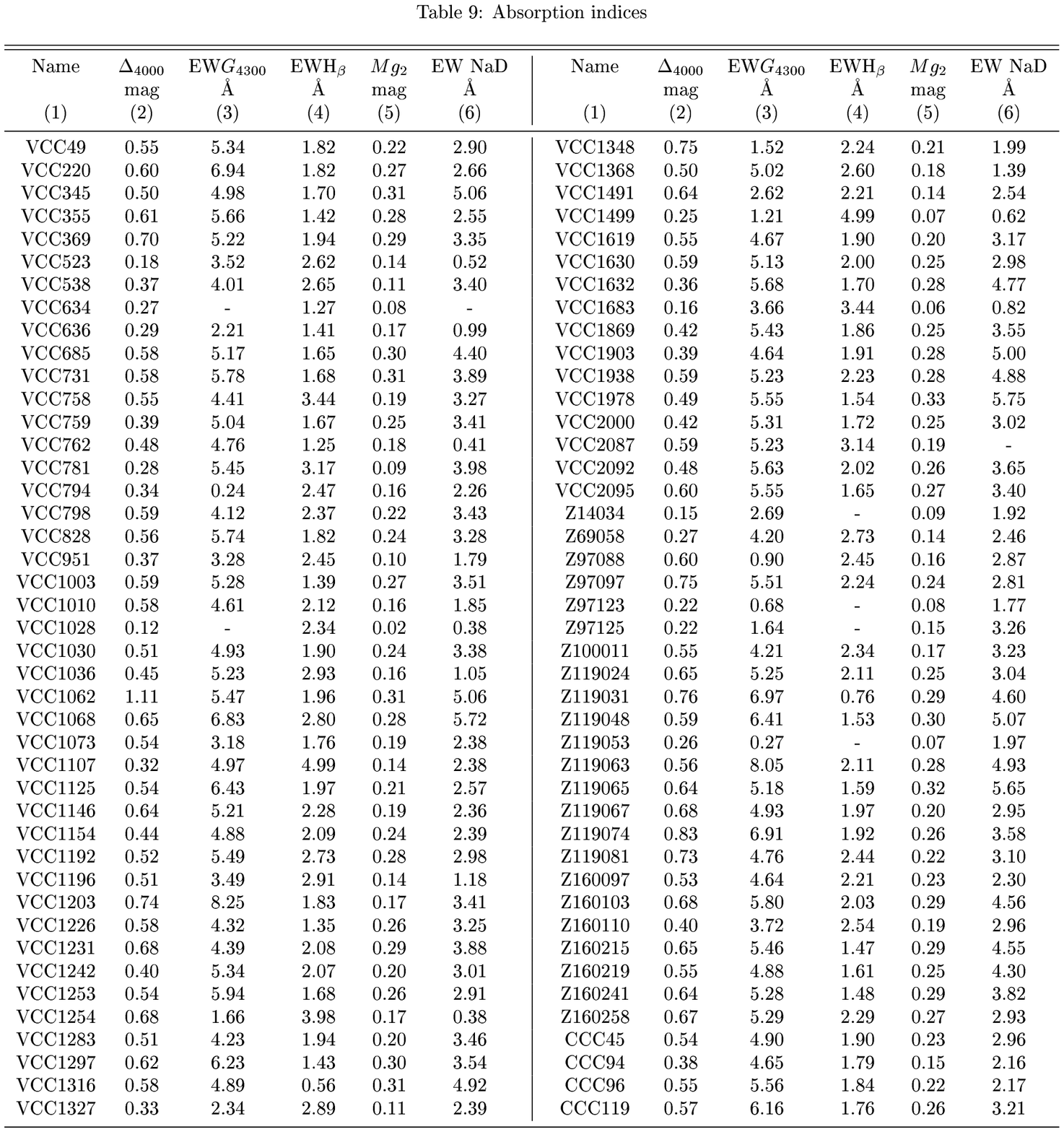,width=16cm,height=23cm}
\end{figure*}

\setcounter{figure}{18}
\begin{figure*}
\psfig{figure=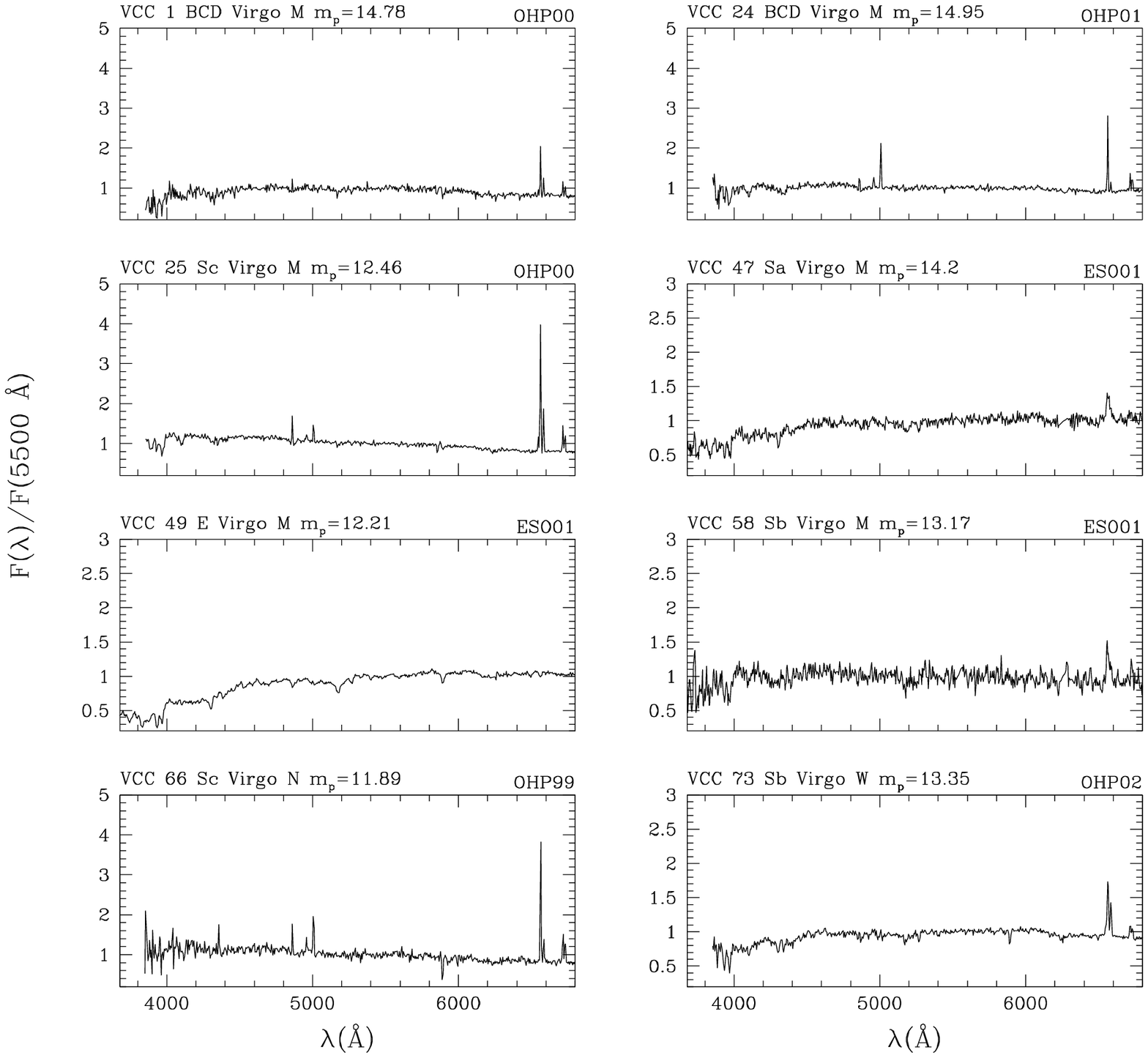,width=15cm,height=11cm}
\psfig{figure=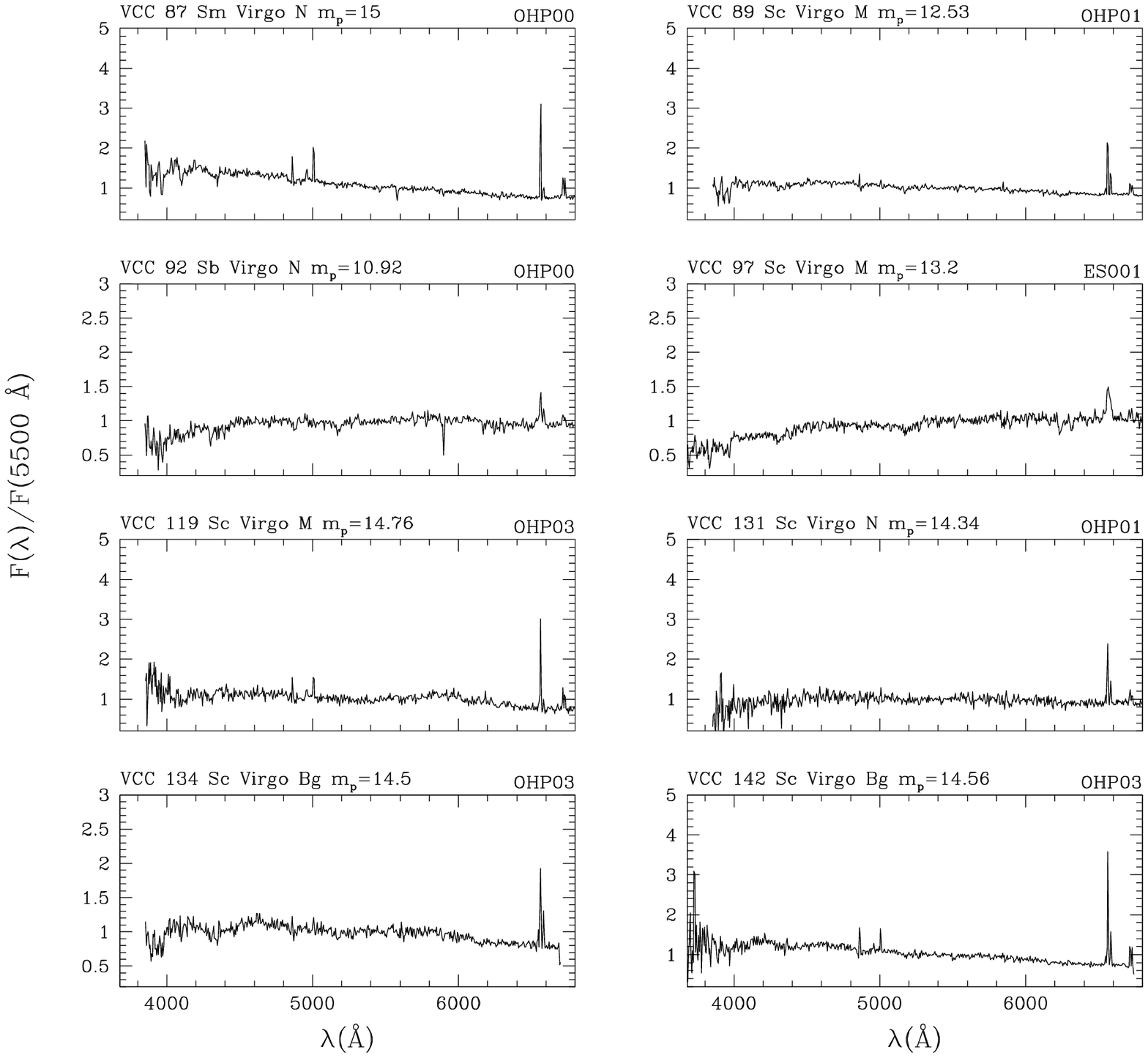,width=15cm,height=11cm}
\caption{The observed spectra. The galaxy identification, morphological type,  
membership, photographic magnitude and observing run are labeled on each panel.
This is one page sample. 
The entire figure is avaible at URL http://goldmine.mib.infn.it/papers/vccspec\_2.html}\label{spec}
\end{figure*}

\end{document}